\documentclass[aps,pre,twocolumn,floatfix,showpacs]{revtex4-1}

\usepackage{graphicx}
\usepackage{amssymb}
\usepackage{color}
\usepackage{amsmath}
\usepackage{amsbsy}
\usepackage{amsmath}
\usepackage{epsfig}
\newcommand\be{\begin{eqnarray}}
\newcommand\ee{\end{eqnarray}}
\newcommand\beq{\begin{equation}}
\newcommand\eeq{\end{equation}}

\begin{document}

\title{Eigenstate Gibbs Ensemble in Integrable Quantum Systems}

\author{Sourav Nandy$^1$, Arnab Sen$^1$, Arnab Das$^1$, Abhishek Dhar$^2$}

\affiliation{$^1$Department of Theoretical Physics, Indian Association
for the Cultivation of Science, Jadavpur, Kolkata 700032, India. \\
$^2$International Centre for Theoretical Sciences, TIFR, Shivakote Village, Hesaraghatta Hobli, Bengaluru 560089, India.}
\begin{abstract}
	
The Eigenstate Thermalization Hypothesis implies that for 
a thermodynamically large system in one of its eigenstates, 
the reduced density matrix describing any finite subsystem is determined solely by a set of {\it relevant} 
conserved quantities.  In a generic system, only the energy plays that role and hence eigenstates appear locally thermal.  
Integrable systems, on the other hand, possess an extensive number of such conserved quantities and hence the reduced density 
matrix requires specification of an infinite number of parameters (Generalized Gibbs Ensemble). 
However, here we show by unbiased statistical sampling of the individual eigenstates with a 
given finite energy density, that the local description of an overwhelming majority of these states of even 
such an integrable system is actually Gibbs-like, i.e. requires only the energy density of the eigenstate. 
Rare eigenstates that cannot be represented by the Gibbs ensemble can also be sampled efficiently by our method and their 
local properties are then shown to be described by appropriately {\it truncated} Generalized Gibbs Ensembles. 
We further show that the presence of these rare eigenstates differentiates the model from the generic (non-integrable) case and 
leads to the system being described by a Generalized Gibbs Ensemble at long time 
under a unitary dynamics following a sudden quench, 
even when the initial state is a Gibbs-like eigenstate of the 
pre-quench Hamiltonian.

\end{abstract}

\date{\today}

\maketitle

\section{Introduction}
\label{intro}

The question of thermalization, i.e., whether or not a closed many-body quantum system 
can act as a heat-bath for its own subsystems when the rest of the system is much bigger, has 
remained an open issue of fundamental importance since the inception of
quantum mechanics. 
The basis for  classical statistical mechanics is 
the hypothesis of equal a priori probability (EAP), which states that all
microstates with equal energy are equally likely to occur during the time
evolution of 
a closed generic (interacting) system  (see, e.g.,~\cite{Landau_Lifshitz}). This gives a possible justification of the use of the microcanonical ensemble. 
On the other hand, a quantum many-body system, prepared in an energy eigenstate, remains in the same energy state.  In this case, EAP is extended 
to the level of single many-body eigenstates resulting in the Eigenstate
Thermalization Hypothesis (ETH)~\cite{Srednicki,Deutsch, Rigol_nature, Luca_etal_Review_ETH}. 
ETH implies that even if a generic many-body system 
is kept in one of its eigenstates, its (local) subsystems are provided with enough quantum fluctuations 
by the rest of the system, so that they can be described by the most general (unbiased) 
ensemble compatible with the conservation of energy of the total system.
Thus suppose that the total system, described by a Hamiltonian $H$,  
is in an eigenstate $|\psi \rangle$, and is described by the corresponding density matrix $\rho = |\psi \rangle \langle \psi|$. 
Then it is expected that the reduced density matrix of the subsystem $\mathcal{S}$, $\rho_{\mathcal{S}} = \mathrm{Tr}_{\bar{\mathcal{S}}} \rho$, obtained by integrating out its complement $\bar{\mathcal{S}}$, should be described by an 
effective density matrix of the form
$\mathrm{Tr}_{\bar{\mathcal{S}}} \rho_{\mathrm{GE}}~,$
where $\rho_{\mathrm{GE}} =\frac{1}{\mathcal{Z}}\exp^{-\beta H}$,  with ${\cal Z}$ being the relevant normalization constant (partition function), and the parameter $\beta$ (inverse temperature)
is fixed solely by requiring that $\rho_{\mathrm{GE}}$ gives an energy density which equals that of the eigenstate.
ETH was implicit in the foundations of quantum statistical mechanics 
(see, e.g.,~\cite{Neumann,Landau_Lifshitz}). 

However, there are important classes of systems e.g., those which can be mapped to non-interacting 
degrees of freedom (see, e.g., ~\cite{BKC-Book, Subir-Book}), where there are infinitely many (of the order of the size of the system) {\it relevant} 
conserved quantities that restricts the statistical distributions of the subsystems. 
If one used an entropy maximisation principle (as in \cite{Jaynes}), 
these conserved quantities are to be treated in the same footing as energy, and that implies a ``generalized" Gibbs 
ensemble (GGE) for the subsystem,  which is characterized by as many parameters as there are conserved quantities 
~\cite{Jaynes}. Extension of this to the eigenstate level implies a restricted (generalized) ETH: 
such systems  would  effectively be described by a reduced
density matrix of the form
$\mathrm{Tr}_{\bar{\mathcal{S}}} \rho_{\mathrm{GGE}}~,$
where $\rho_{\mathrm{GGE}} = \exp{\left[-\sum_{i}^{{\mathcal N}_{L}} \lambda_{i} \hat{\cal{I}}_{i}\right]}/{\cal Z}$, where $\hat{{\cal I}}_{i}$ denotes the
relevant integrals of motion, $\lambda_{i}$ are their corresponding Lagrange multipliers and ${\cal N}_{L}$ is proportional to the system size~\cite{Jaynes, Rigol_Prl,Cassidy_etal,Caux_Essler_PRL}. 
An important question here is whether {\it all} the integrals of motion 
$\hat{\cal{I}}_{i}$ are necessary to describe the properties of a finite subsystem. 
This idea of equilibrium statistical mechanics has been extended to describe the asymptotic synchronized states of periodically driven non-interacting systems
(or those mappable to it) using periodic Gibbs' ensemble~\cite{AAR-PRL}, hence the question is not necessarily limited to the domain of equilibrium statistical mechanics.

Another related approach to thermalization is to start from a pure state,
usually the ground state of a local (pre-quench) Hamiltonian, that is not 
an eigenstate of the system's final (post-quench) Hamiltonian, and let it
evolve 
in time under the resulting unitary
dynamics~\cite{Rigol_nature,Rigol_Prl,Calabrese_Cardy_PRL,Kollath_Lauchli_Altman_PRL,Kris-Rev}. 
If the system can act as its own reservoir, as ETH implies, then the long-time
evolved state can also be described by a thermal density matrix as far as
local operators 
are concerned. However, if the evolution of the state is due to an integrable
Hamiltonian, the 
long time behavior of local operators should instead be again described by a
GGE (and not GE) 
which respects the extensive number of conservation laws forced by the unitary
dynamics 
of the (post-quench) Hamiltonian. Whether the infinite amount of information
 regarding all the conserved quantities $\hat{\cal{I}}_{i}$ is really
 necessary 
to understand local properties is again an important issue in describing steady states that eventually arise from such dynamics. 

In this work, we consider the finite energy density eigenstates of the
transverse field Ising model (TFIM) in one dimension ($1$D) and study the
reduced density matrices (RDMs) and local correlation functions in subsystems
of $l$ consecutive spins by performing an unbiased sampling of the individual
eigenstates in chains of linear dimension $L$ (with $L$ ranging upto $10^5$
spins). By doing a careful finite-size scaling, we find that the RDMs of a
{\it typical} finite energy density eigenstate approaches the standard GE form
(and not a GGE) determined only by the energy density of the eigenstate for $l
\ll L$, but not for finite $l/L$, as $L \rightarrow \infty$. This is inspite
of the integrable nature of the model and is because the densities of all the
additional ``local'' conserved quantities approach their ``thermal''
values as $L \rightarrow \infty$, and so the corresponding Lagrange
multipliers vanish. This provides an explicit example of {\it weak}
ETH~\cite{Canonical_typicality_PRL,PRL_Lauchli_rarestates} where typical (but
not all) energy eigenstates appear thermal when local correlation functions
are probed.  We note that such a weak ETH scenario has been recently
numerically demonstrated in a different kind of (Bethe integrable) spin model~\cite{Alba_PRB_ETH} 
and a Bose gas~\cite{Ueda_etal_weakETH} in one dimension. However, only the
infinite temperature ensemble was considered in Ref.~\onlinecite{Alba_PRB_ETH}, while we have no such restriction on the average energy density of the sampled eigenstates. 
Moreover these studies obtained eigenstates using Bethe ansatz, and so were limited to small system sizes. Furthermore, we also consider the local properties of the rare eigenstates where the effects of the other integrals of motion (apart from the Hamiltonian) becomes apparent. The presence of (rare) eigenstates which do not follow a GE locally in the thermodynamic limit is a consequence of the integrability of the model, since such states are believed to be absent in a generic system (for numerical tests of the same, see Refs.~\onlinecite{SantosRigolPRE,Kim_Ikeda_Huse,Masud_etal_ETH}). The fraction of such rare eigenstates shrinks to zero in the thermodynamic limit but these can also be sampled efficiently by our method and their local properties are then shown to be described by RDMs that approach appropriate ``truncated'' GGEs as $L \rightarrow \infty$ and $l \ll L$ where only a {\it few} ($\mathcal{O}(1)$) integrals of motion need to be retained for an accurate description for a majority of such states.

Furthermore, we also consider a sudden quench of the magnetic field in the
$1$D TFIM where the initial state is not the ground state of the pre-quench
Hamiltonian (see also Ref.~\onlinecite{Calabrese_quench_excited_state}) but instead a typical finite energy density eigenstate, and study the nature of the steady state obtained at asymptotically large times. We show that even though the initial (pure) state is locally thermal, the final state needs a full GGE description for its local properties. The behaviour of the Lagrange multipliers in the GGE however has important differences compared to a quench starting from the ground state of the pre-quench Hamiltonian~\cite{Fagotti_Essler_PRB}, which we point out here.

The rest of the paper is arranged in the following manner. In Sec.~\ref{basics}, we review some results relevant for our work and set the notations for the rest of the paper. In Sec.~\ref{MC}, we describe our numerical procedure for sampling any given finite energy density eigenstates of the $1$D TFIM in chains of size $L$. The behaviour of the typical eigenstates is described in Sec.~\ref{results_typical}, and we consider the rare eigenstates that requires a GGE description in Sec.~\ref{results_rare}. In Sec.~\ref{quench}, we obtain an analytic expression for the GGE which describes the steady state after a quench, where the initial state is a typical finite energy density eigenstate. Finally, we summarize our results and conclude in Sec.~\ref{conclude}.

\section{$1$D TFIM: some preliminaries}
\label{basics}

The $1$D TFIM is defined by the following Hamiltonian:
\be
H =-\sum_{j=1}^L (g \sigma_j^x + \sigma_j^z \sigma_{j+1}^z)
\label{tfim}
\ee
where $\sigma^{x,y,z}$ are the Pauli operators and the external magnetic field equals $g$. We further impose periodic boundary condition ($\sigma_{L+1}^\alpha = \sigma_1^\alpha$ where $\alpha=x,y,z$) with $L$ being even. The ground state of this model is ferromagnetic when $-1<g<1$ and paramagnetic otherwise, with continuous quantum critical points at $g = \pm1$~\cite{Subir-Book}. 

This model can be solved exactly for any finite $L$ using a well-known mapping of the spins to {\it spinless fermions} (Jordan-Wigner transformation) (e.g. see Ref.~\onlinecite{BKC-Book,Subir-Book}): 
\be
\sigma_n^x &=& 1-2c^{\dagger}_n c_n \nonumber \\ 
\sigma_n^z &=& -(c_n+c_n^{\dagger})\prod_{m=1}^{n-1}(1-2c^{\dagger}_m c_m) 
\label{jw}
\ee
 From Eq.~\ref{jw}, the vacuum state of the $c$ fermions, which we denote by $|0\rangle$, corresponds to $\sigma^x=+1$ for all sites. Writing $H$ (Eqn.~\ref{tfim}) in terms of these fermions, we obtain (after omitting constant terms)
\be
H &=& 2g\sum_{j=1}^L c^{\dagger}_j c_j -\sum_{j=1}^{L-1} \left(c^{\dagger}_jc_{j+1}+c^\dagger_{j}c^\dagger_{j+1} + \mathrm{h.c.} \right) \nonumber \\
  &+& (-1)^{N_F}[c_L^\dagger c_1+ c_L^\dagger c_1^\dagger + \mathrm{h.c.} ]
\label{fermionH} 
\ee
The sign of the boundary term depends on whether the total number $N_F$ of the
$c$ fermions is odd or even. If $N_F$ is odd, periodic boundary conditions on
the fermions is required ($c_{L+1} = c_1$), whereas for $N_F$ even,
antiperiodic boundary condition is imposed ($c_{L+1} = -c_1$). Since the
Hamiltonian conserves fermion parity, these sectors do not mix and we restrict ourselves to even $N_F$ for the rest of this paper.

To diagonalize the Hamiltonian, we go to momentum space and accordingly define
\be
c_k &=& \frac{\exp(i\pi/4)}{\sqrt{L}}\sum_x \exp(-i k x)c_x 
\ee 
where $k = 2\pi m/L$ with $m = -(L-1)/2, \cdots, -1/2,1/2, \cdots , (L-1)/2$. Re-writing $H$ in terms of $c_k,c_k^{\dagger}$, we get $H = \sum_{k>0} H_k$ where 
\be 
H_k &=& 2(g-\cos(k))[c^{\dagger}_k c_k - c_{-k}c^{\dagger}_{-k}] \nonumber \\
&+& 2 \sin(k) [c_{-k}c_k + c^{\dagger}_k c^{\dagger}_{-k}]~.
\label{hk2}
\ee
This Hamiltonian connects the vacuum (of the $c$ fermions) $|0 \rangle$ with $|k,-k \rangle =c_k^{\dagger}c^{\dagger}_{-k}|0 \rangle$, and $|k \rangle = c_k^{\dagger} |0 \rangle$ with $|-k \rangle = c_{-k}^{\dagger}|0 \rangle$. 

We further restrict ourselves to the parity invariant states (PIS) in which
all the positive and negative momentum modes are populated with the same
weights. All the eigenstates $|\psi \rangle$ of the TFIM at a magnetic field
strength $g$ which are also PIS can then be written in the form
\be 
|\psi \rangle &=& \otimes_{k>0}|\psi_k \rangle \nonumber \\
|\psi_k \rangle &=& U_{kn}(g)c^{\dagger}_k c^{\dagger}_{-k} |0 \rangle + V_{kn}(g)|0 \rangle 
\label{producteigenstates}
\ee
 where $(U_{kn}(g),V_{kn}(g))$ can only have either of the two forms shown below at each $k$ to be an eigenstate: 
\be
(U_{k0}(g),V_{k0}(g)) &=& \left(-\sin \left(\frac{\theta^g_k}{2} \right),\cos\left(\frac{\theta^g_k}{2} \right) \right) \nonumber \\
(U_{k1}(g),V_{k1}(g)) &=& \left(-\cos\left(\frac{\theta^g_k}{2} \right),-\sin\left(\frac{\theta^g_k}{2} \right)\right) \nonumber \\
\sin(\theta^g_k) &=& \frac{\sin(k)}{\sqrt{(g-\cos(k))^2+(\sin(k))^2}}
\label{eigenstates1}
\ee
These eigenstates can be equivalently represented by strings with either $0$ or $1$ at each $k>0$, which we denote by the label $n_k$, where $0 (1)$ refers to $(U_{k0(1)}(g),V_{k0(1)}(g))$. The total energy of such an eigenstate is given by 
\be
E&=&\sum_{k>0}\epsilon_k(g) (2n_k-1) \nonumber \\
\epsilon_k(g) &=& 2\sqrt{(g-\cos(k))^2+(\sin(k))^2}
\label{energyexp1}
\ee
These states represent $2^{L/2}$ of the $2^L$ eigenstates of the TFIM
(including its ground state) in a chain of length $L$ and we will focus exclusively on these states in this study. A quantum quench (by suddenly changing the magnetic field $g$) in which the initial state is such an eigenstate of the TFIM continues to be a PIS (though not an eigenstate of the post-quench Hamiltonian) under the unitary dynamics.  

For completeness, note that $H_k$ in Eq.~\ref{hk2} can be easily diagonalized through a Bogoliubov rotation with an angle $\theta_k^g/2$ (with $\theta^g_k$ as defined in Eq.~\ref{eigenstates1}) to give 
\be
H = \sum_{k>0} \epsilon_k(g) (\mathcal{A}_k^\dagger \mathcal{A}_k +\mathcal{A}_{-k}^\dagger \mathcal{A}_{-k} -1)
\label{energyexp2}
\ee
 with $\mathcal{A}_k=V^*_{k0}(g)c_k - U_{-k0}(g)c^{\dagger}_{-k}$ denoting the
 Bogoluibov fermion operator at momentum $k$. Thus $n_k$ equals $\mathcal{A}_k^\dagger \mathcal{A}_k = 0 (1)$ and represents such an unoccupied (occupied) single-particle level at momentum $k$. Since we are considering parity invariant eigenstates, the Bogoluibov fermions at $k$ and $-k$ are always (un)occupied in pairs giving Eq.~\ref{energyexp1} from Eq.~\ref{energyexp2}.

\subsection{Local properties of individual eigenstates}
\label{localRDMs}

\subsubsection{Generalized Gibbs ensemble}
\label{integrals}
To write down the GGE description for the individual eigenstates (Eq.~\ref{producteigenstates}) in the TFIM or for the steady state obtained after a quantum quench, we need to specify the extensive number of integrals of motion $\hat{\cal{I}}_{i}$ present in the model. From the (non-local) mapping of the spins to free fermions using the Jordan-Wigner transformation as discussed in the previous section, it is clear that the average occupation number of the Bogoluibov fermion at each momentum $k$, i.e. $n_k = \mathcal{A}_k^\dagger \mathcal{A}_k$, is a conserved quantity and the number of such conserved quantities scales extensively with $L$. For the case of quantum quenches, the following GGE construction~\cite{Rigol_Prl,Cassidy_etal} has been shown to provide the correct description for properties of the steady state of the system: 
\be
 \rho_{\mathrm{GGE}}=\frac{1}{\mathcal{Z}} \exp(-\sum_{k}\lambda_k n_k)
\label{GGE_standard}
\ee 
where the Lagrange multiplier $\lambda_k$ is defined as
\be 
\lambda_k = \log \left(\frac{1-\langle n_k \rangle}{\langle n_k \rangle} \right)
\label{lambdak}
\ee
with $\langle n_k \rangle = \langle \psi|\mathcal{A}_k^\dagger \mathcal{A}_k|\psi \rangle $, where $\mathcal{A}_k^\dagger \mathcal{A}_k$ refers to the Bogoluibov fermion occupation of the post-quench Hamiltonian in case of the quantum quench.   

This form of the GGE, however, does not make it clear as to which conserved
quantities need to be retained and which can be ignored when describing local
properties of the system, since the occupation numbers $n_k$ are {\it
  non-local} in real space. Moreover, these conservations do not possess
corresponding local densities, unlike the Hamiltonian. Another problem with
this form of $\rho_{\mathrm{GGE}}$ arises when considering exact eigenstates
of the TFIM, and not the steady state following a quench, since there the
corresponding Lagrange multipliers $\lambda_k$ are not defined microscopically
as each $n_k$ can only be $0$ or $1$. 

An equivalent representation of $\rho_{\mathrm{GGE}}$ was recently constructed by Fagotti and Essler for the TFIM~\cite{Fagotti_Essler_PRB} where only the local (in space) conservations $I^{\pm}_n$ (where $n$ is a non-negative integer) present in the model were considered for constructing the GGE. Each such $I^{\pm}_n$ involves $n+2$ neighboring spins but can be written in a straightforward manner in terms of the occupations numbers $n_k$~\cite{Fagotti_Essler_PRB} as
\be
I_n^{+}&=& \sum_k \cos(nk) \epsilon_k(g) n_k \nonumber \\ 
I_n^{-} &=& -\sum_k 2 \sin[(n+1)k] n_k 
\label{In_def}
\ee
Again, it is implicit here that $n_k$ in the definition of $I_n^{\pm}$ refers
to the average Bogoluibov fermion occupation of the post-quench Hamiltonian in
the case of a quantum quench. 

The GGE can now be defined in terms of these local integrals of motion as 
\be
\rho_{\mathrm{GGE}} = \frac{1}{\mathcal{Z}}\exp \left(-\sum_{n=0}^{(L/2)-1} \sum_{\sigma=\pm}[\lambda_n^{\sigma}I_n^{\sigma}] \right) 
\label{GGE_FE}
\ee
where the Lagrange multipliers $\lambda_n^{\sigma}$ are fixed by the conditions:

\be
 \mathrm{Tr}[\rho_{GGE}I_n^\sigma] = {\langle \psi|I_n^\sigma|\psi \rangle}~.
\ee

This representation of the RDMs serves as the ideal starting point for the issues that we address here. Firstly, as was shown in Ref.~\onlinecite{Fagotti_Essler_PRB} for the case of quantum quenches in the TFIM, the properties of local subsystems with $l$ consecutive spins in the final steady state can be understood by only considering the $y$ most local conservation laws, i.e., 
\be
\rho^{(y)}_{GGE} = \frac{1}{\mathcal{Z}_y}\exp \left(-\sum_{n=0}^{y-1} \sum_{\sigma=\pm}[\lambda_{n,y}^{\sigma}I_n^{\sigma}] \right)
\label{GGE_y}
\ee 
where $y \sim \mathcal{O}(l)$ gives a very good description of the sub-system properities and including more non-local conservation laws only gives an exponentially small $\exp(-y)$ correction thereafter. Thus, for describing the properties of subsystems of size $l$, $I_n^{\sigma}$ with $n \gg l$ can be completely ignored. We will show later that similar behaviour occurs for the RDMs for $l$ consecutive spins when finite energy density eigenstates of the TFIM are considered, with $y \sim \mathcal{O}(l)$ providing a very good description of the subsystem. Secondly, unlike $\lambda_k$, the Lagrange multipliers $\lambda_{n,y}^{\sigma}$ are well-defined microscopically for the eigenstates.

For the eigenstates which are also PIS (Eqn.~\ref{producteigenstates}), it is easy to see that $I_n^{-}=0$ because $n_k = n_{-k}$ (Eqn.~\ref{In_def}). Thus, we need to only consider the integrals of motion $I_n^{+}$ and will henceforth suppress the index $+$ from both $I_n^{+}$ and $\lambda_n^+$. Also, $I_0$ equals the total energy of the system (shifted such that the ground state has zero energy) and thus the Lagrange multiplier $\lambda_0$ can be identified with the inverse temperature $\beta$. Since both the descriptions of $\rho_{\rm GGE}$ are equivalent (Eqn.~\ref{GGE_standard} and Eqn.~\ref{GGE_FE}), it is possible to transform from $\lambda_k$ to $\lambda_n$ by using:
\be
\lambda_n = \left(\frac{2-\delta_{n,0}}{L}\right) \sum_{k>0} \frac{\lambda_k}{\epsilon_k}\cos(n k).
\label{lambdan}
\ee
\subsubsection{Reduced density matrices and the distance measure}

We proceed in a similar way to Ref.~\onlinecite{Kitaev_PRL_entanglement} to calculate the entanglement of $l$ adjacent spins for the TFIM. The RDMs for any individual eigenstate of the form Eqn.~\ref{producteigenstates} is most simply calculated after expressing that state in terms of the $c$ fermions. Since the transformation between the spins and the fermions is non-local, we {\it cannot} express the RDM of $l$ non-adjacent spins in any simple manner in terms of the fermion correlations involving only the sites within the subsystem. However, if we take $l$ adjacent spins as the subsystem, then all the non-zero spin correlations involving any subset of these $l$ sites for a finite $L$ can be expressed in terms of the fermionic correlation functions at these $l$ sites~\cite{Essler1,Essler2}. This is straightforward to see for correlation functions involving $\sigma^x_n$ only since these are local in terms of the $c$ fermions. Moreover, even for correlations functions involving an even number of $\sigma^z_n$, ($\sigma^z_n$ being non-local in terms of the $c$ fermions, see Eqn.~\ref{jw}), the Jorgan-Wigner strings outside the subsystem cancel and the resulting expression is in terms of the fermions within the subsystem only. Correlations functions with an odd number of $\sigma^z_n$ are zero due to the $\mathcal{Z}_2$ symmetry of the model. The RDM can then be calculated solely by considering the correlation functions of the fermions in the subsystem. Furthermore, since the fermions are non-interacting, all higher point fermionic correlators can be calculated from the two-point correlation functions using Wick's theorem~\cite{Peschel_PRB_freefermions}.

The two-point fermionic correlations can be expressed in terms of two $l \times l$ matrices~\cite{Peschel_PRB_freefermions}, 
${\bf C}$ and ${\bf F}$, whose elements are constructed by knowing
($U_{kn}(g)$, $V_{kn}(g)$) for the eigenstate $|\psi \rangle$
(Eqn.~\ref{producteigenstates} and Eqn.~\ref{eigenstates1}): 

\be 
C_{ij} &=& \langle \psi| c^{\dagger}_{i} c_{j} |\psi \rangle = \frac{2}{L} \sum_{k>0} |U_{kn}(g)|^2 \cos(k (i-j)) \nonumber \\
F_{ij} &=& \langle \psi| c^{\dagger}_{i} c^{\dagger}_{j}|\psi \rangle = \frac{2}{L} \sum_{k >0} U^*_{kn}(g) V_{kn} (g)\sin(k(i-j)) \nonumber \\  
\label{matrices1}
\ee 
where  $i,j$ refer to sites in the subsystem.

The RDM for a block of $l$ adjacent spins may then be written in terms of the $c$ fermions as 
 \be
\rho_\mathcal{S} &=& \frac{1}{Z_\mathcal{S}}\exp(-\mathcal{H}_\mathcal{S}), \nonumber \\
\mathcal{H}_\mathcal{S} &=& \sum_{k=1}^l \mathcal{E}_{k,\mathcal{S}} \eta_{k,\mathcal{S}}^\dagger \eta_{k,\mathcal{S}}
\label{reducedr} \ee
 
where $\mathcal{H}_\mathcal{S}$ denotes its ``entanglement Hamiltonian''
which is diagonal in terms of operators
$\eta_{k,\mathcal{S}},\eta_{k,\mathcal{S}}^\dagger$ that are fermionic operators for single particle
states with energies $\mathcal{E}_{k,\mathcal{S}}$ and linearly related to the
operators $c_i,c^{\dagger}_i$.
 $Z_{\mathcal{S}}$ ensures the
correct normalization $\mathrm{Tr}(\rho_\mathcal{S})=1$. 

Since {\it all}
correlation functions of the subsystem can be expressed in terms of
the quadratic fermionic correlations by using Wick's Theorem here, the entanglement Hamiltonian $\mathcal{H}_\mathcal{S}$, and hence $\rho_\mathcal{S}$, is
fully determined by the condition that it gives the {\it right} quadratic
correlation functions $C_{ij}$ and $F_{ij}$ for the sites that
belong to the subsystem~\cite{Peschel_PRB_freefermions}. Calculating $\rho_{\mathcal{S}}$ thus requires only the eigenvectors and eigenvalues of the $2l \times 2l$ matrix $\mathcal{C}$ defined as 
\be \quad
\begin{pmatrix}
\mathbf{I-C} & \mathbf{F^{\dagger}} \\
\mathbf{F} & \mathbf{C}
\end{pmatrix}
\ee

 Particularly, the entanglement entropy of the subsystem $S_{ent}(l)$ only requires the eigenvalues: 
\be
S_{ent}(l) &=& -\mathrm{Tr}(\rho_\mathcal{S} \log \rho_\mathcal{S}) \nonumber \\
&=& -\sum_{k=1}^{2l}p_k \log(p_k)
\ee 
where $p_k$ denotes the eigenvalue of the $\mathcal{C}$ matrix. 

We now define a distance measure for the RDMs in an eigenstate $|\psi \rangle$
to quantify how well these operators are described by the truncated GGEs based
on a few local integrals of motion. Since all the local conservations $I_n$
are quadratic in the $c$ fermions (Eqn.~\ref{In_def}), one can simply define the distance measure using the correlation matrices $\mathcal{C}(l)$ and $\mathcal{C}_{\mathrm{GGE}}^{(y)}(l)$~\cite{AS-KS}, where the latter is calculated assuming the density matrix of the full system to be $\rho_{\mathrm{GGE}}^{(y)}$ (Eqn.~\ref{GGE_y}). We use the standard trace distance between these two matrices to define the distance measure $\mathcal{D}(\mathcal{C}(l),\mathcal{C}^{(y)}_{\mathrm{GGE}} (l))$ as 
\be
 \frac{1}{2l} \mathrm{Tr} \sqrt{(\mathcal{C}^{(y)}_{\mathrm{GGE}} (l)-\mathcal{C}(l))^\dagger (\mathcal{C}^{(y)}_{\mathrm{GGE}} (l)-\mathcal{C}(l))}
\ee
Note that $0 \leq \mathcal{D}(\mathcal{C}(l),\mathcal{C}^{(y)}_{\mathrm{GGE}}
(l)) \leq 1$ and is identically zero only when
$\mathcal{C}^{(y)}_{\mathrm{GGE}} (l)=\mathcal{C}(l)$. When
$\mathcal{D}(\mathcal{C}(l),\mathcal{C}^{(y)}_{\mathrm{GGE}} (l))=0$, it
implies that all the (non-zero) correlation functions $\langle \psi |
\mathcal{O} | \psi \rangle$, where $\mathcal{O}$ is defined using any subset
of the $l$ spins in the subsystem, 
coincides with the values obtained from the corresponding truncated GGE.

\section{Algorithm for sampling eigenstates}
\label{MC}
For a large chain of size $L$, since there are $2^{L/2}$ eigenstates that are parity invariant, it is not possible to extract the local properties for each individual state in a numerical calculation. Instead, we use an unbiased sampling procedure which we detail below, to extract individual eigenstates from a microcanonical ensemble with the mean value of the energy density $e=\langle \psi|H|\psi \rangle /L$ being equal to the ``target'' energy density $e_T$ and the fluctuations around the mean $\Delta e \rightarrow 0$ as $L \rightarrow \infty$. Our sampling is based on the standard algorithm for performing a microcanonical Monte-Carlo (MC) simulation introduced by Creutz~\cite{Creutz_PRL}, where an extra degree of freedom, which we call ``demon'', travels throughout the system exchanging energy with it, and changing the dynamical variables as a result. 

In the context of the TFIM, we can think of the demon traveling in $k$ space, and attempting to update the Bogoluibov fermion occupations $n_k (=0(1))$ which fully define the eigenstate (Eqn.~\ref{producteigenstates}). In detail, a momentum $k$ from the allowed positive momenta at system size $L$ is chosen at random. Upon reaching $k$, the demon attempts to flip the variable $n_k$ from $0(1)$ to $1(0)$. If this move lowers the energy of the system $E=\sum_{k>0} 2\epsilon_k(g) n_k$, this energy is then given to the demon and the flip is accepted. The demon energy, which we denote by $E_D$, is then updated to $E_{D^{'}}$ as follows 
\be
E_D \rightarrow E_{D^{'}} = E_D + E - E^{'}
\label{demonenergy}  
\ee    
where $E^{'}$ is the new energy of the system. Note that the total energy of
the system and the demon remains conserved in this process. Similarly, if the
system's energy is increased by the flip, the demon supplies that required
energy and its own energy is decreased accordingly. However, to keep the demon
from running off with all the energy, we restrict $E_D \geq 0$ and so only
those flips are accepted for which $E_{D^{'}} \geq 0$, otherwise the flip is
rejected. This Monte-Carlo (MC) procedure thus generates an unbiased random
walk in the space of configurations with $E+E_D = E_T$ since the transition
$(E_D,E) \rightarrow (E_{D^{'}}, E^{'})$ and its reverse are allowed with
equal probability. The mean energy density of the sampled energy eigenstates
during the MC can be tuned to a required target energy density $e_T$ by
starting with an initial demon energy $E_D=0$ and choosing an initial
eigenstate with the appropriate energy density $E_T/L$. 
The width in the energy densities of the sampled eigenstates $\Delta e
\rightarrow 0$ 
as $L \rightarrow \infty$ since $E_D \ll E$ as $L \gg 1$. We define one
 Monte-Carlo step (MCS) as $L/2$ flip attempts by the demon, and use 
the first $10^4$ MCS as warm-up so that the memory of the initial eigenstate 
choice is lost, and then use the next $10^6$ MCS for measurements of the individual properties of these sampled eigenstates.

\section{Properies of typical eigenstates}
\label{results_typical}
 To understand the local properties of the {\it typical} eigenstates from a
 microcanonical ensemble with a desired mean energy density $e_T$ at a
 magnetic field strength $g$, we sample such states using our MC and measure
 $\langle \psi |\sigma^x| \psi \rangle$, $\langle \psi |I_1| \psi \rangle /L$, the distance measure $\mathcal{D}(\mathcal{C}(l),\mathcal{C}^{(y)}_{\mathrm{GGE}} (l))$ (all of which may be readily calculated in the $c$ fermion representation) with different choices of truncated GGEs for subsystems of $l$ adjacent spins and the entanglement entropy $S_{\mathrm{ent}}(l)$ of such a block for each of the generated eigenstate. 

\begin{figure}[htb]
{\includegraphics[width=\hsize]{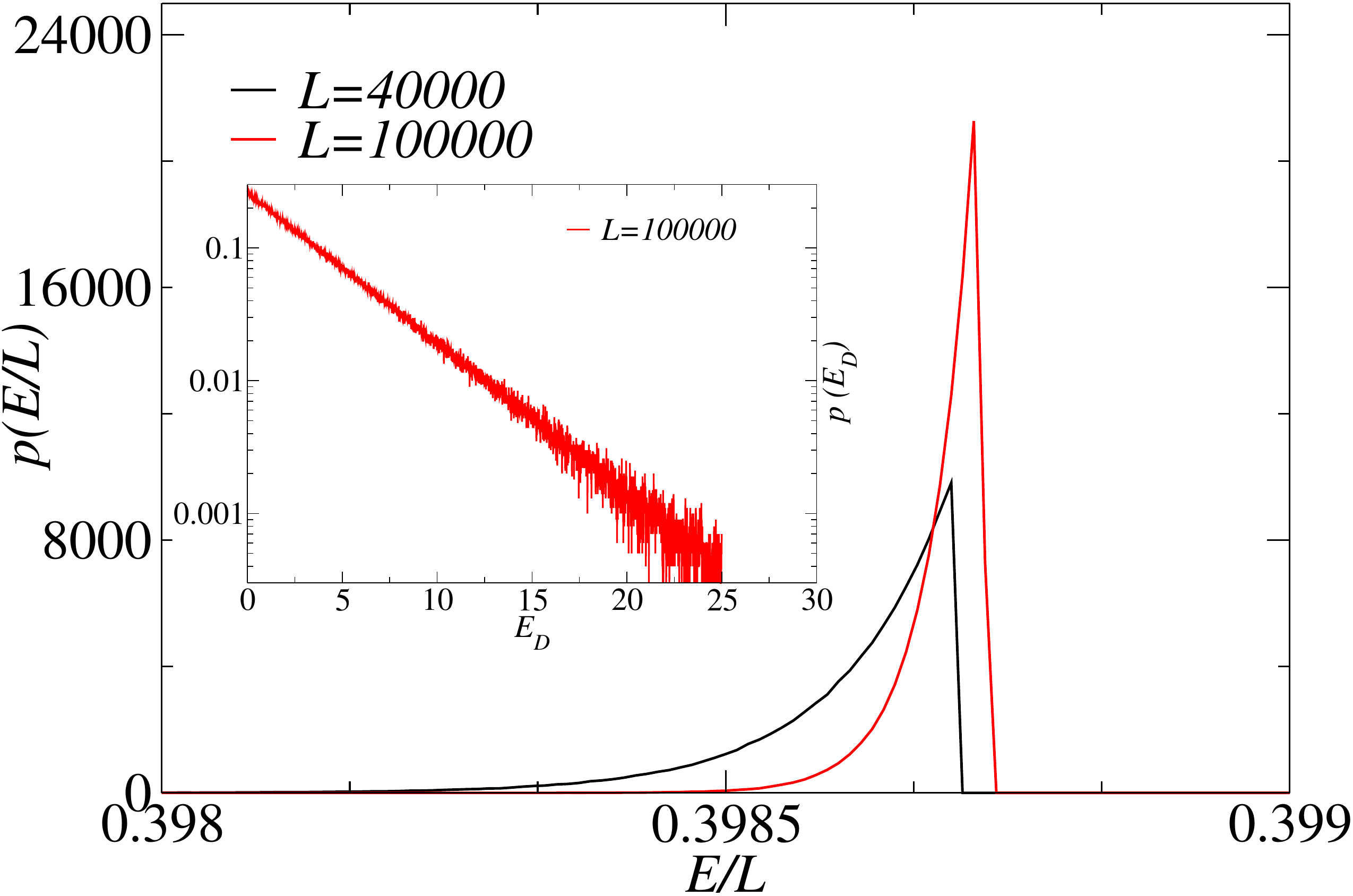}}
\caption{ Probability density of $E/L$ for the sampled eigenstates at coupling
  $g=2$ and target energy density of $e_T = 0.3986$. For the chain sizes used,
  the range of the sampled $E/L$ is very small and mimics a Microcanonical
  ensemble. The inset shows the behaviour of the demon energy $E_D$ during the
  sampling procedure.   
\label{paperfig1}}
\end{figure}

Here, we show the results of the MC for $g=2$ with a mean energy density of
$e_T=0.3986$ (within error bars). The average demon energy $\langle E_D
\rangle$ is finite and equals $3.84$ (Fig.~\ref{paperfig1}, inset). 
Firstly, we see that for large chain sizes, the sampled eigenstates 
have an energy density $E/L$ which has a very narrow spread that 
rapidly shrinks to zero with increasing $L$ (see Fig.~\ref{paperfig1}), 
thus leading to an unbiased sampling of eigenstates from the microcanonical
ensemble.

\begin{figure}[htb]
{\includegraphics[width=\hsize]{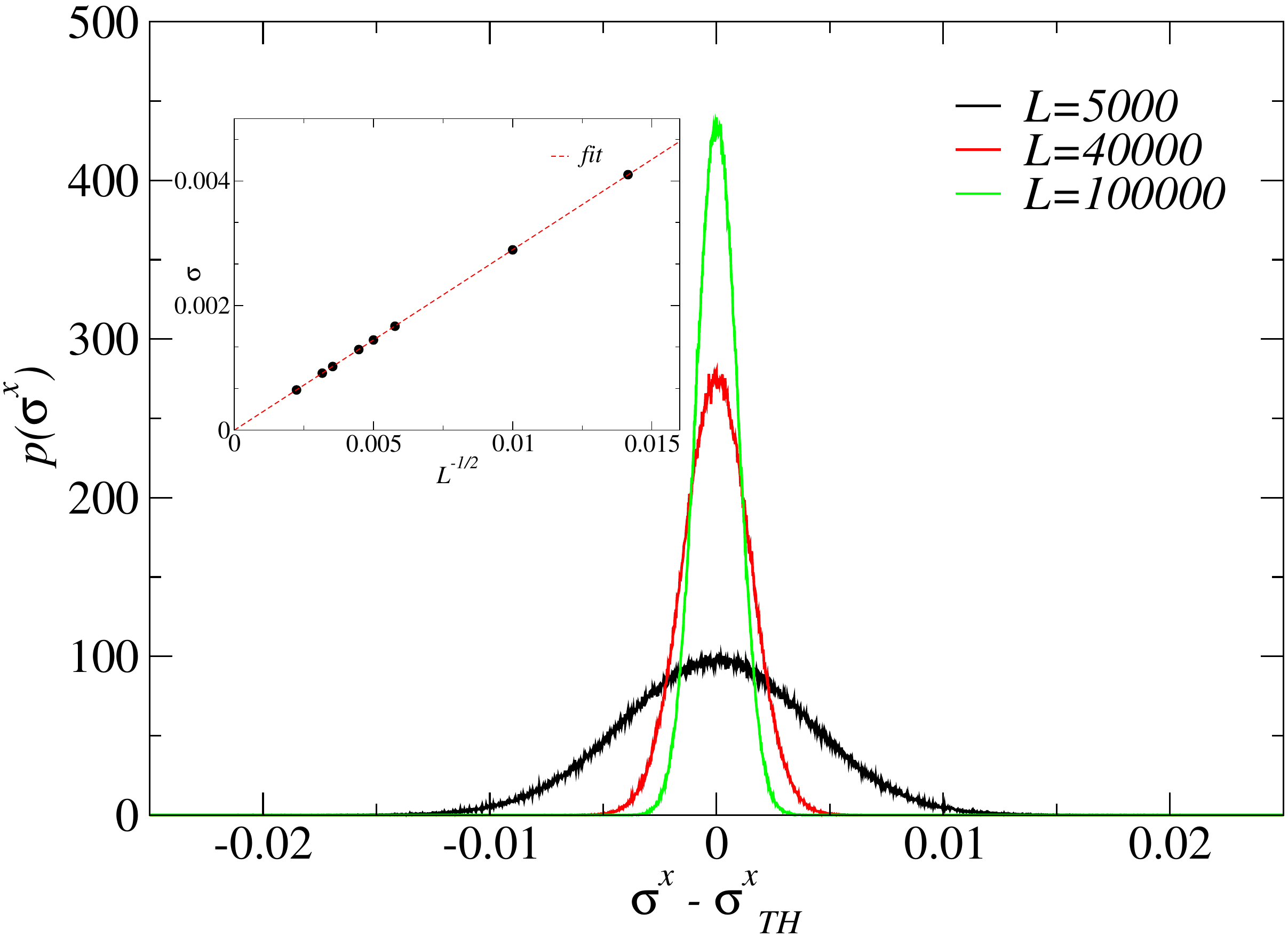}} \\
{\includegraphics[width=\hsize]{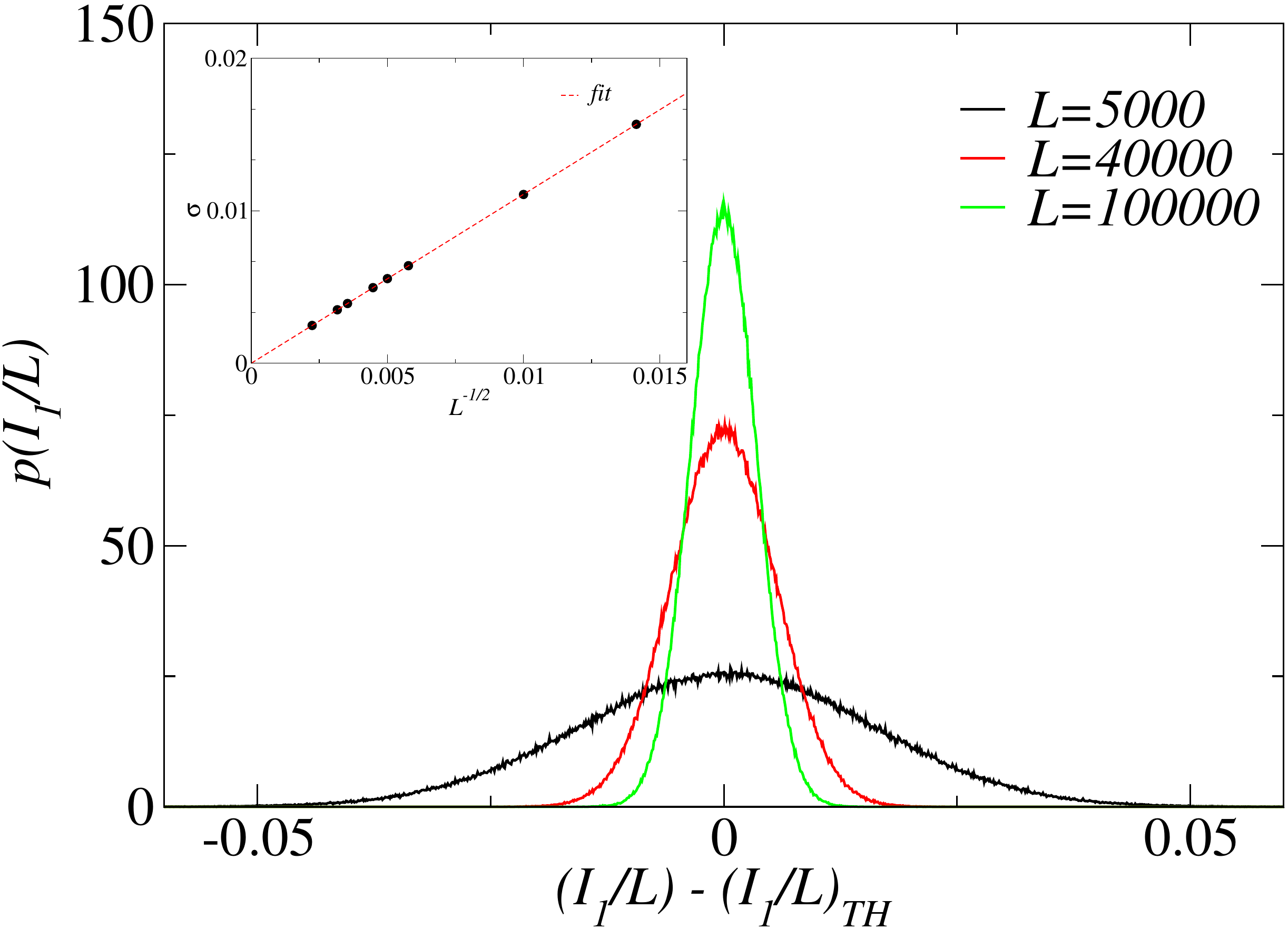}}
\caption{Probability density generated from the sampled values of
  $\langle \psi |\sigma^x| \psi \rangle$ (Top panel) and $\langle \psi |I_1| \psi
  \rangle /L$ (Bottom panel) 
for various $L$ at $g=2$ from within a Microcanonical ensemble with average
energy density of $e_T = 0.3986$. 
$\sigma^x_{\rm{TH}}$ and $(I_1/L)_{\rm{TH}}$ denote the corresponding thermal
values at a 
finite $\beta$ fixed only by the average energy density $e_T$ in the
thermodynamic limit. 
The insets of both the figures show that the standard deviation $\sigma$ decreases as $L^{-1/2}$.   
\label{paperfig2}}
\end{figure}

We show the probability densities of the sampled values of $\langle \psi
|\sigma^x| \psi \rangle$ and $\langle \psi |I_1| \psi \rangle /L$  
obtained from the MC in
Fig.~\ref{paperfig2}. The sampled values have a Gaussian
distribution whose mean depends only on $e_T$ for a given $g$ and standard
deviation that decays to zero as $L^{-1/2}$ (insets of
Fig.~\ref{paperfig2}). This numerical evidence strongly suggets that, in the
thermodynamic limit, the local properties for any {\it typical} eigenstate of
the TFIM (the atypical states contribute to the tails of the distributions
becoming increasingly rare with increasing system size during the MC) depends
{\it only} on the energy density $e_T$. Then, the natural ensemble to get the
local properties correctly as $L \rightarrow \infty$ is the GE where the
inverse temperature $\beta$ is calculated from the mean energy density $e_T$,
and is $\beta = 0.2604$ in this case. 

This is indeed what is observed when the mean values of $\langle \psi
|\sigma^x| \psi \rangle$ (Fig.~\ref{paperfig2}, top panel) and $\langle \psi |I_1| \psi
\rangle/L$ (Fig.~\ref{paperfig2}, bottom panel) are
calculated from the sampled eigenstates. Since the
width around the mean shrinks to zero when $L \rightarrow 0$, typical
eigenstates have the corresponding thermal values for $\langle \psi |\sigma^x|
\psi \rangle$ and $\langle \psi |I_1| \psi \rangle / L$ in this limit. Indeed,
normal distribution of the fluctuations about the mean thermal value and the
$L^{-1/2}$ scaling of the standard deviation 
was also observed in free models~\cite{PRL_Lauchli_rarestates} and 
in quantities studied in Ref.~\onlinecite{Alba_PRB_ETH}, and may be 
a generic feature of many observables in typical eigenstates of 
integrable models at finite sizes.

\begin{figure}[htb]
{\includegraphics[width=\hsize]{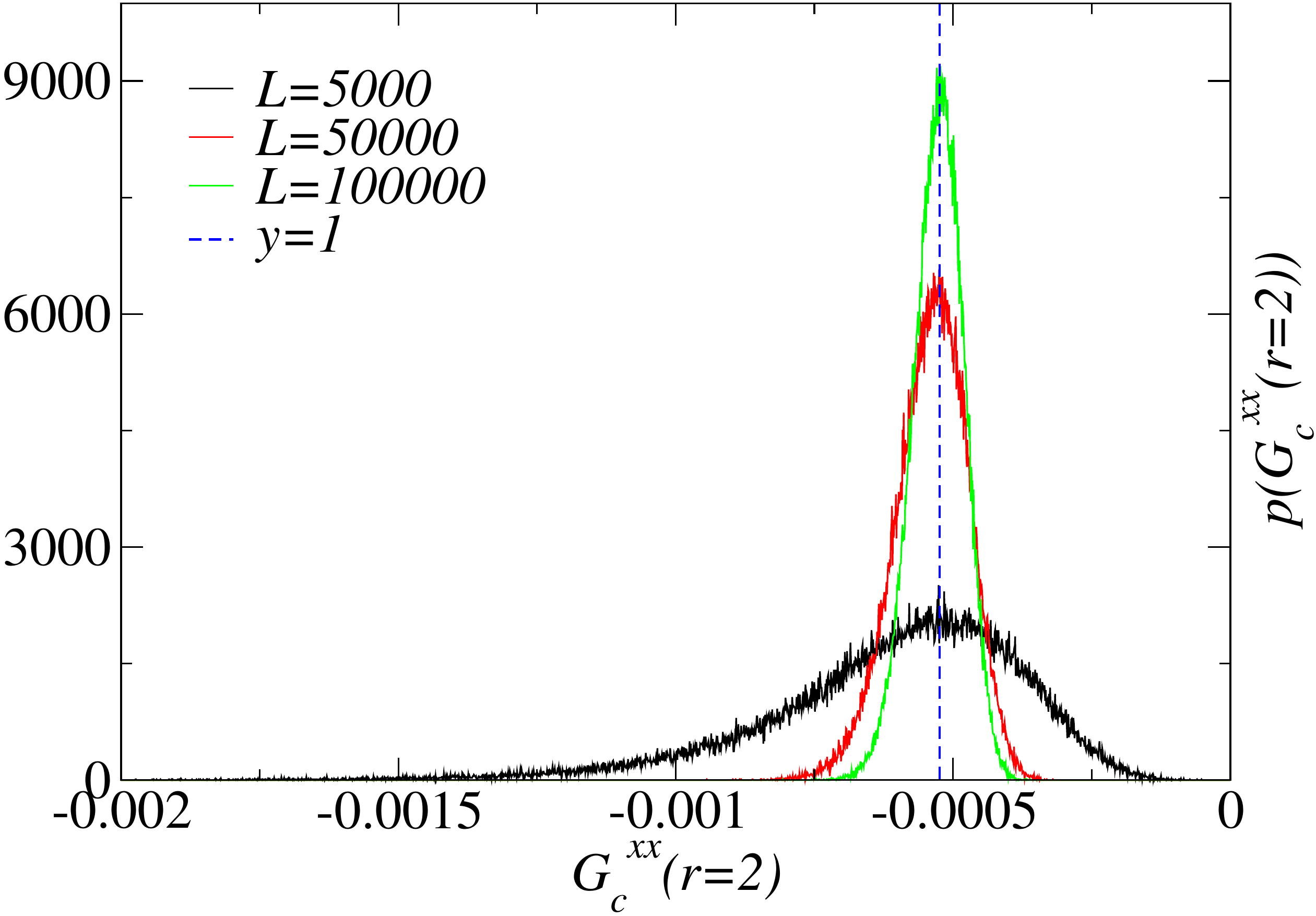}}
\caption{ Probability density generated from the sampled values of $G_c^{xx}(r) = \langle \psi |\sigma^x(r+1) \sigma^x(1)| \psi \rangle- \langle \psi| \sigma^x| \psi \rangle^2$ for $r=2$ from within a Microcanonical shell with energy density $e_T = 0.3986$. The probability density is asymmetric about the thermal mean value but shrinks to it when $L \rightarrow \infty$.    
\label{paperfig3}}
\end{figure}

It is useful to note here that that not all local operator expectation values in these typical eigenstates are normally distributed about the thermal mean value at finite chain size $L$. E.g., we show the behaviour of the connected correlation function $G_c^{xx}(r) = \langle \psi |\sigma^x(r+1) \sigma^x(1)| \psi \rangle- \langle \psi| \sigma^x| \psi \rangle^2$ for $r=2$ in Fig.~\ref{paperfig3}, where the distribution function is clearly asymmetric (and not Gaussian about the corresponding thermal mean value) but shrinks to the thermal value again as $L \rightarrow \infty$. 
 
\begin{figure}[htb]
{\includegraphics[width=\hsize]{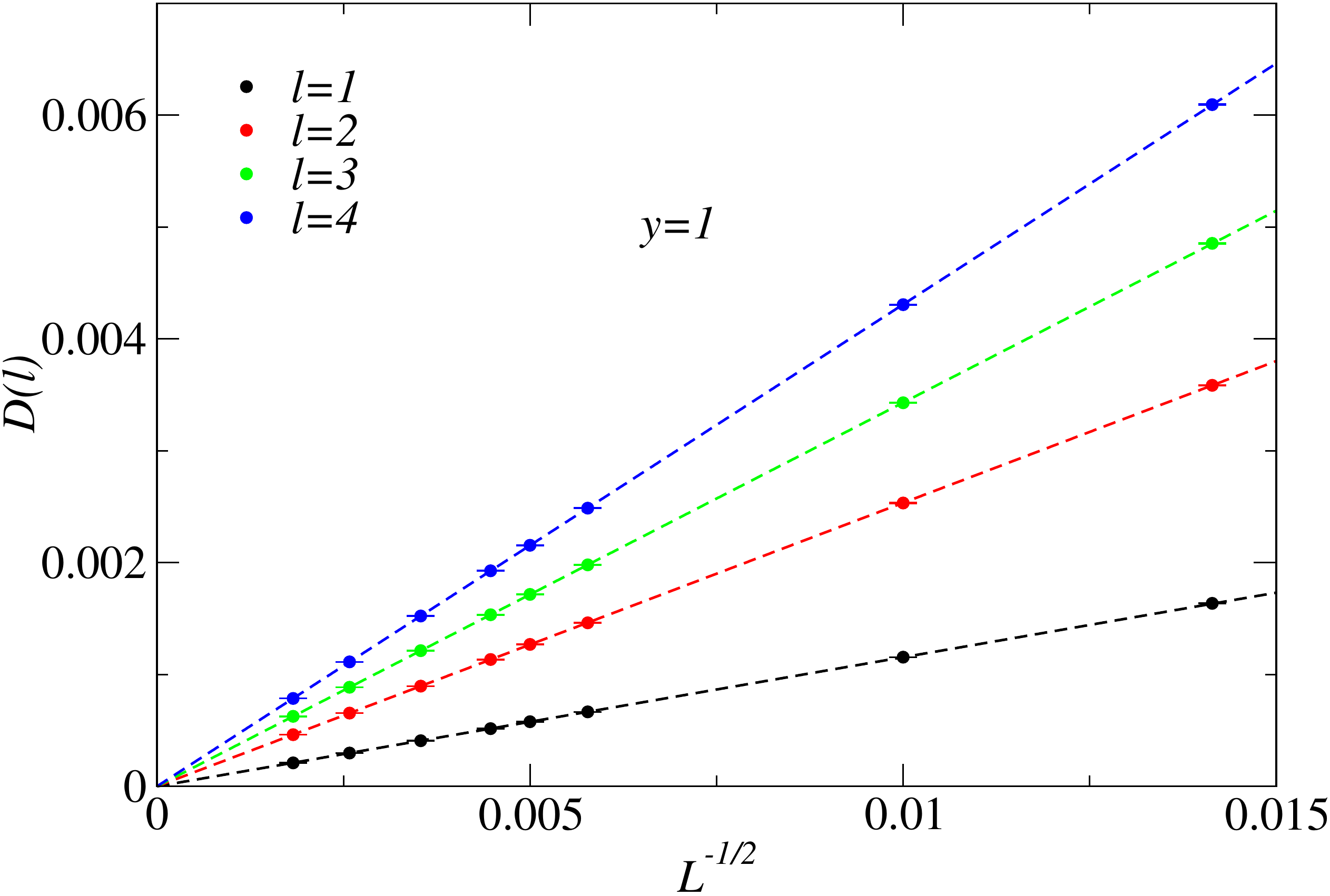}} \\
{\includegraphics[width=\hsize]{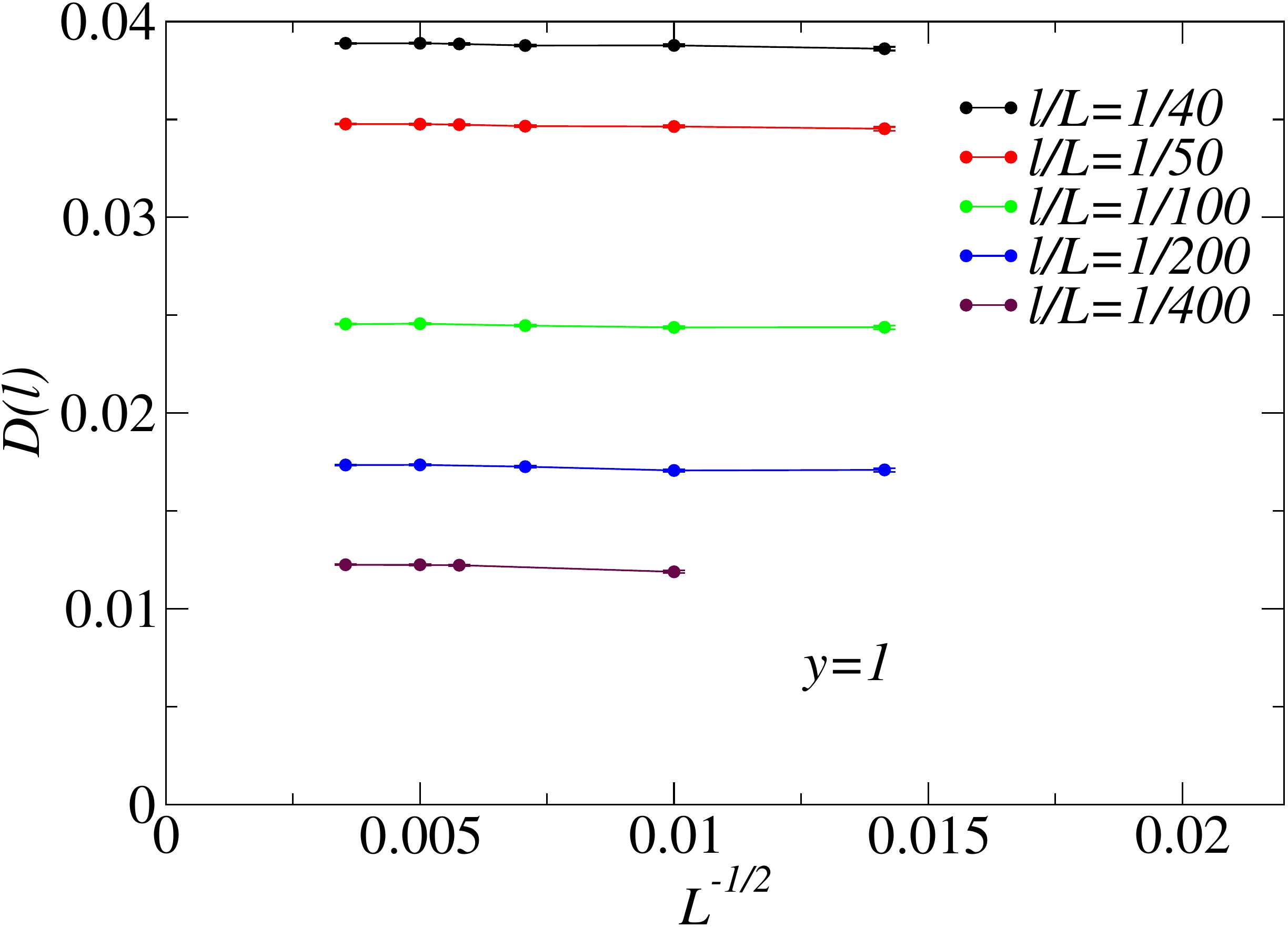}} 
\caption{(Top panel) The distance measure  $\mathcal{D}(C(l),C^{(1)}_{\rm{GGE}}(l)) \rightarrow 0$ as $L \rightarrow \infty$ when $l \ll L$ implying that the local RDMs are thermal. This is however not the case when subsystems with finite $l/L$ are considered (Bottom panel).    
\label{paperfig4}}
\end{figure}

To show unambigiously that the equivalence to GE holds at the level of the RDMs, which implies that 
\be
\rho_{\mathcal{S}} = \mathrm{Tr}_{\bar{\mathcal{S}}} |\psi \rangle \langle \psi| = \mathrm{Tr}_{\bar{\mathcal{S}}} \rho_{\mathrm{GE}} 
\ee
 where $\rho_{\mathrm{GE}} = \frac{1}{\mathcal{Z}} \exp(-\beta H)$ for a typical eigenstate $|\psi \rangle$ when $L \rightarrow \infty$ as long as the subsystem is local (i.e. $l \ll L$), we consider the behaviour of
the average $\mathcal{D}(C(l),C^{(1)}_{\rm{GGE}}(l))$ (where the truncated GGE
with $y=1$ coincides with GE, and we have used the inverse temperature $\beta
= 0.2604$ which is fixed to give the the correct average energy density $e_T$
of the sampled eigenstates) and see that the distance measure itself goes to
zero for the typical states as $L \rightarrow \infty$ (see
Fig.~\ref{paperfig4}, top panel), again as $L^{-1/2}$ at large $L$ (we have
also verified this for bigger subsystems till $l \leq 100$). This implies that
{\it all} typical eigenstates are locally described by a GE in the
thermodynamic limit. We also see that if subsystems with finite $l/L$ are
considered, then the distance measure does not go to zero as $L \rightarrow
\infty$ even when $l/L$ is very small (Fig.~\ref{paperfig4}, bottom panel). This is because global operators which involve spins at a spatial seperation $l \sim \mathcal{O}(L)$ cannot be described by a thermal reduced density matrix as the rest of the system cannot then act as a bath for the subsystem. 

\begin{figure}[htb]
{\includegraphics[width=\hsize]{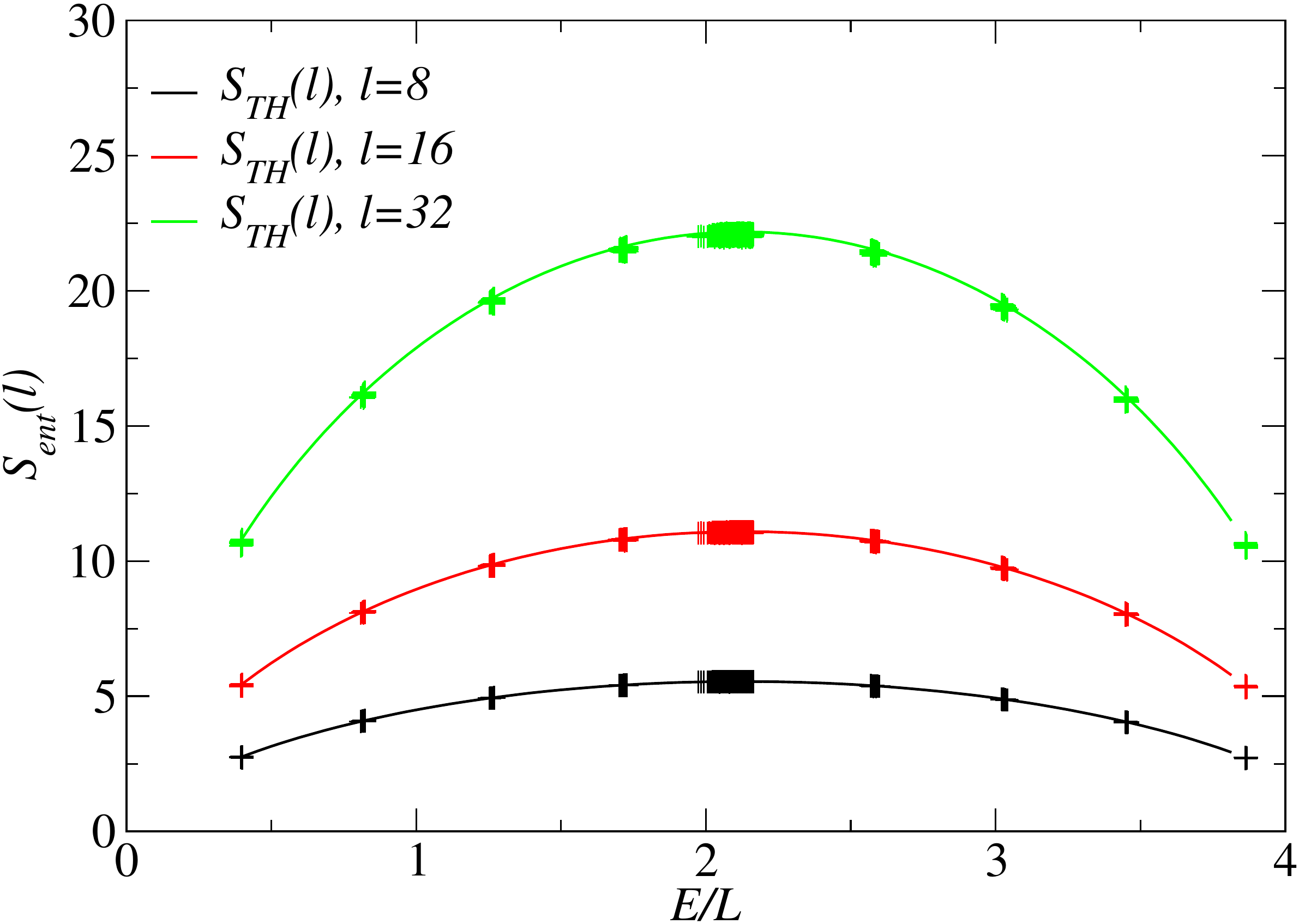}} 
\caption{ The entanglement entropy of small subsystems of size $l$, denoted by $S_{\rm{ent}}(l)$, obtained from the typical eigenstates sampled at different energy densities $E/L$ at coupling $g=2$ for a chain size of $L=5000$. These match very well with the corresponding thermal entropy $S_{\rm{TH}}(l)$ obtained from the average energy density of the sampled eigenstates.    
\label{paperfig5}}
\end{figure}

Finally, we show evidence that this feature, of typical eigenstates locally behave as if they are thermal, holds at all values of energy density $e_T$ for the coupling $g=2$. With our MC, we can also access eigenstates with a {\it negative} values of $\beta$ (i.e., eigenstates which lie {\it above} the middle of the spectrum) by restricting the demon energy to be $E_D \leq 0$ (instead of $E_D \geq 0$) in the MC and these continue to be described by the corresponding GEs. To demonstrate the local thermal behaviour, we calculate the entanglement entropy $S_{\rm{ent}}(l)$ directly from the sampled eigenstates and see that these agree very well with the corresponding thermal value of the entropy $S_{\rm{TH}}(l)$ assuming a GE for the full system (see Fig.~\ref{paperfig5}). Since the spectrum of the TFIM is bounded, the entanglement shows a non-monotonic behaviour with varying energy density. We have further checked that typical eigenstates at other values of the magnetic field $g$ also behave thermally as far as local properties are concerned. Since we are considering a one-dimensional model here, such eigenstates are always paramagnetic (i.e. $\langle \psi| \sigma^z| \psi \rangle = 0$) in the thermodynamic limit for any finite energy density irrespective of the value of $g$.  

The thermal behaviour of the local observables in the typical eigenstates of the TFIM in the thermodynamic limit can be related to an analogous behaviour of the Bogoluibov fermion occupation $n_k$ which determine the densities of {\it all} the (local) conserved quantites of the model (Eqn.~\ref{In_def}). When $L \rightarrow \infty$, we get 
\be
\frac{I_n}{L} = \frac{1}{\pi} \int_0^\pi \cos(nk) \epsilon(k) n_c(k)  
\label{coarsegrainednk}
\ee
where the momentum $k$ becomes a continuous variable and $n_c(k) \in [0,1]$ represents the average occupation of the Bogoluibov fermions at momentum $k$.

For free fermions, it was demonstrated in Ref.~\onlinecite{ETH_freefermions}
(see also Refs.~\onlinecite{RRPSingh_freefermions,SDSarma_freefermions}) that
if a ``coarse-grained'' occupation number $n_c(k)$, defined through some
suitable averaging procedure of the microscopic variables $n_k$ in a shell of
(infinitesimal) width $\delta k$ around $k$ is considered, then the most
probable form of $n_c(k)$ appears thermal (i.e. the Fermi-Dirac distribution
for free fermions) by the standard entropy maximization argument. Clearly,
many different microscopic realizations of $n_k = 0,1$ can give the same
``coarse-grained'' $n_c(k)$ in the thermodynamic limit, which explains the
resulting thermal values of the densities, $I_n/L$, for the typical
eigenstates as $L \rightarrow \infty$ (Fig.~\ref{paperfig2}). For a finite system size of $L$, there are $\mathcal{O}(\delta k L)$ momentum modes in a shell of width $\delta k$ around $k$, and hence fluctuations of $\mathcal{O}(1/\sqrt{L})$ that are normally distributed around the most probable $n_c(k)$ can be expected for typical eigenstates by the Central Limit Theorem. This explains the normal distribution of $\langle \psi | \sigma^x |\psi \rangle$ and $\langle \psi | I_1 |\psi \rangle/L$ about the corresponding thermal values at finite $L$ (Fig.~\ref{paperfig2}) since these quantities depend linearly on the fermion occupation.

\section{Sampling atypical eigenstates}
\label{results_rare}
The demon algorithm can be easily generalized to generate atypical eigenstates
from within the Microcanonical ensemble that do not satisfy GE. These states
are characterized by {\it athermal} values of the densities $I_n/L$ and there
is again a large number of such eigenstates ($\mathcal{O}(e^L)$) for a chain
of size $L$. We adapt our algorithm to sample energy eigenstates from within a
truncated {\it generalized} Microcanonical ensemble defined by $(E/L, I_1/L,
\cdots)$ where the densities of the other integrals of motion $(I_1/L,\cdots)$
are set to be significantly different from their corresponding thermal values
in the thermodynamic limit. 
Such eigenstates can clearly not be described by a GE locally. Here, we
discuss our 
results for {\it typical} eigenstates from within the simplest (truncated) 
generalized Microcanonical ensemble $(E/L,I_1/L)$ and the extension to other cases is immediate.

\begin{figure}[htb]
{\includegraphics[width=\hsize]{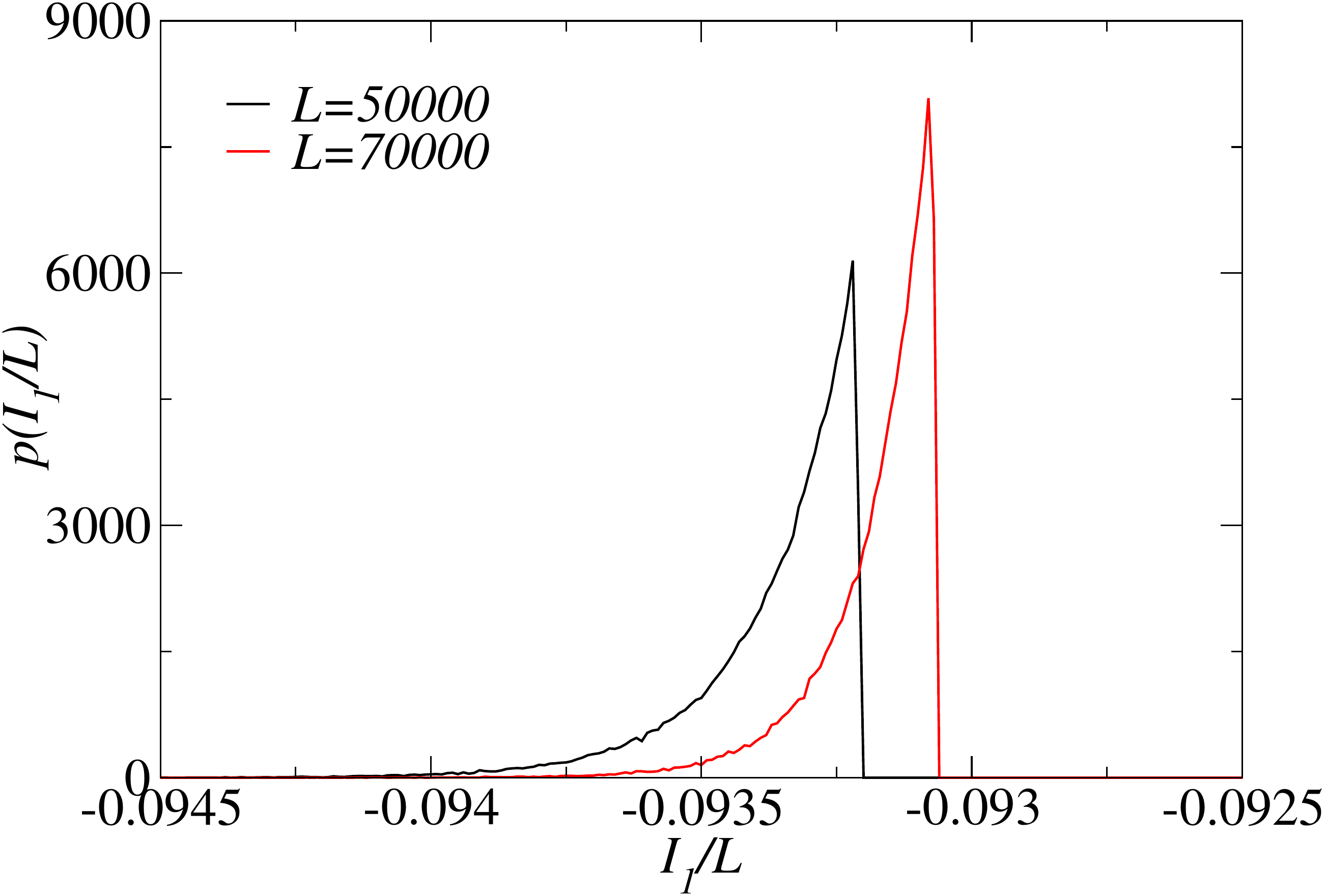}}
\caption{ Probability density of $(I_1/L)$ for the sampled eigenstates when
  the demon has both $E_D$ and $(I_1)_D$ as its properties at the same target
  energy density $e_T = 0.3986$ at $g=2$ as in Fig.~\ref{paperfig1} 
but with very different $I_1/L$ compared to the corresponding thermal value of $(I_1/L)_{\rm TH} = 0.075$.   
\label{paperfig6}}
\end{figure}

 For sampling such rare eigenstates, we now endow the demon with two properties $E_D$ and $(I_1)_D$. The demon again visits a momentum $k$ randomly from the allowed positive momenta at system size $L$ and attempts to flip the variable $n_k$ from $0(1)$ to $1(0)$. The demon variables $E_D$ and $(I_1)_D$ are simultaneously updated as
\be
E_D &\rightarrow& E_{D^{'}} = E_D + E - E^{'} \nonumber \\
(I_1)_D &\rightarrow& (I_1)_{D^{'}} = (I_1)_D + I_1 - I_1^{'}
\ee  
where $E = \sum_{k>0}2\epsilon_k n_k$ and $I_1 = \sum_{k>0}2\epsilon_k n_k
\cos(k)$ as defined earlier, and $E^{'}$ and $I_1^{'}$ are the correspondingly
values after the flip. We further restrict the demon to have $E_D \geq 0$ and
$(I_1)_D \geq 0$ at all times and only those flips which satisfy these
conditions simultaneously are accepted, otherwise the flip attempt is aborted
and another $k$ is chosen at random. We choose the initial seed eigenstate
with appropriate values of $E_T$ and $(I_1)_T$ and initialize $E_D,
(I_1)_D=0$. Since the MC conserves $E+E_D=E_T$ and $I_1+(I_1)_D = (I_1)_T$ and
in a large system, since $E_D \ll E$ and $(I_1)_D \ll I_1$, we therefore only
sample eigenstates with a fixed energy per site {\it and} a fixed density
$(I_1)/L$ when $L \rightarrow \infty$. Here, we show sampling of eigenstates
at $g=2$ with the
 same $E/L = 0.3986$ as in the previous section but 
now with a very atypical value of $(I_1)/L = -0.093$ (Fig.~\ref{paperfig6}) 
which is far from the corresponding thermal value of $(I_1/L)_{\rm{TH}} =+0.075$ given the energy density.

\begin{figure}[htb]
{\includegraphics[width=\hsize]{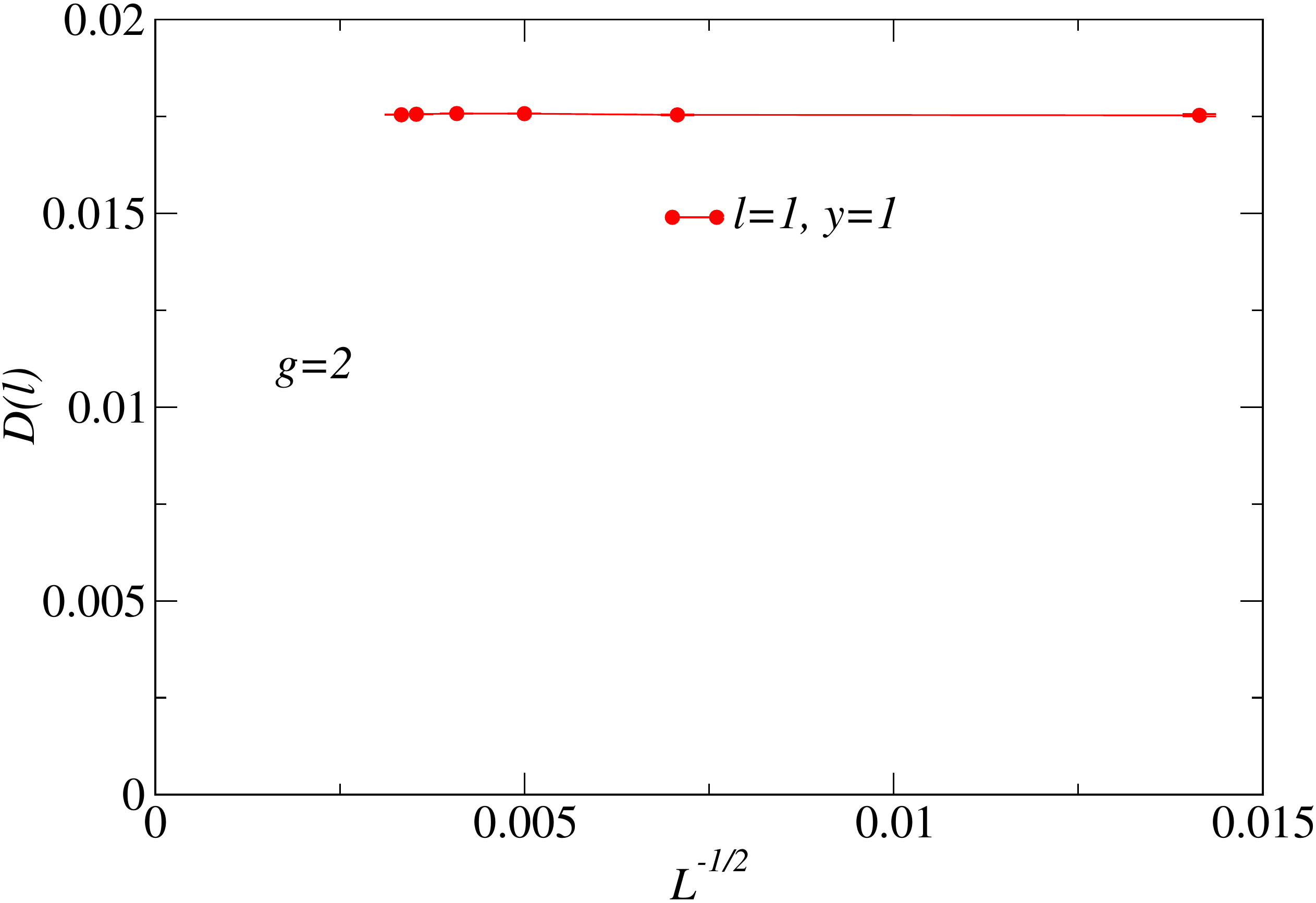}} \\
{\includegraphics[width=\hsize]{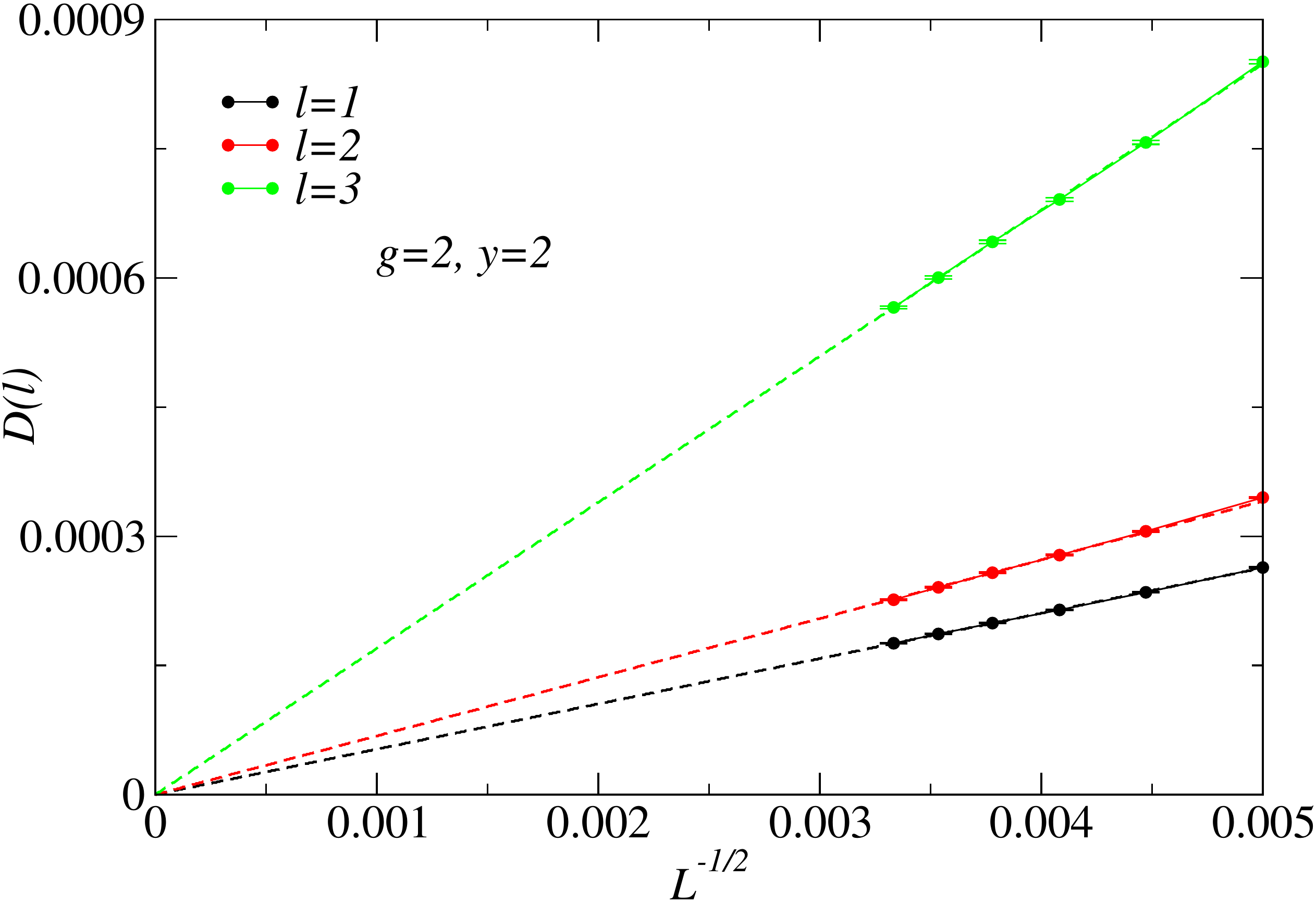}}
\caption{The local properties of {\it typical} eigenstates with $E/L =0.3986$
  and  $(I_1)/L = -0.093$ at $g=2$ (which is very different from
  $(I_1/L)_{\rm{TH}} = +0.075$) are not described by a GE in the thermodynamic
  limit (Top panel), but by a $y=2$ truncated GGE with $\beta = 0.296$ and
  $\lambda_1 = 0.129$ (Bottom panel).
\label{paperfig7}}
\end{figure}

For the typical eigenstates generated in this generalized Microcanonical
ensemble, we clearly see that the RDMs {{\it cannot}} be described with
corresponding GEs, unlike in the previous case (Fig.~\ref{paperfig7}, top
panel) since now the distance measure does not go to zero with $y=1$. However,
the truncated GGE with $y=2$ (i.e. with the athermal density of $I_1$ taken
into account through $\lambda_1$), which gives $\beta = 0.296$ and $\lambda_1
= 0.129$, exactly describes the local properties of these sampled eigenstates
in the thermodynamic limit when $l \ll L$ (Fig.~\ref{paperfig7}, bottom panel). 

Using the basic definition of entropy $S=\log(\Omega)$, where $\Omega$ denotes
the number of eigenstates that share the same local properties, we thus see
that the probability of encountering an eigenstate which is described by a
$y=2$ ensemble characterized by $(E/L,I_1/L)$ (where $I_1/L$ is athermal)
versus an eigenstate described by a GE characterized by $E/L$ alone equals
$\exp[-L\{s(E/L)-s(E/L,I_1/L)\}]$, 
where $s(E/L)$ is the entropy density of the system at inverse temperature
$\beta$ fixed by $E/L$ and $s(E/L,I_1/L)$ is the corresponding entropy density
for the system described by a $y=2$ truncated GGE, with $(\beta,\lambda_1)$
fixed jointly by $(E/L,I_1/L)$. 
This quantifies why such athermal eigenstates are ``rare'' states in the TFIM, even though their number still scales exponentially with system size.     

Similarly, we have numerically verified that a {\it typical} energy eigenstate
from other generalized Microcanonical ensembles with the first $m+1$ local
conservations laws specified is completely characterized (as far as all local
properties are concerned) by a suitable truncated GGE with only $y=m+1$ as $L
\rightarrow \infty$. This behaviour in the thermodynamic limit can be argued
from the corresponding most probable distribution of the coarse-grained (in
momentum space) Bogoluibov fermions occupations $n_c(k)$
(Eqn.~\ref{coarsegrainednk}) by extending the arguments of Ref.~\onlinecite{ETH_freefermions}, taking into account the additional conservation laws which specify the generalized Microcanonical ensembles in terms of the Bogoluibov fermion occupations $n_k$.    

\subsection{Truncated GGE for an arbitrary eigenstate}
\label{sec:truncatedgge}

\begin{figure}[htb]
{\includegraphics[width=\hsize]{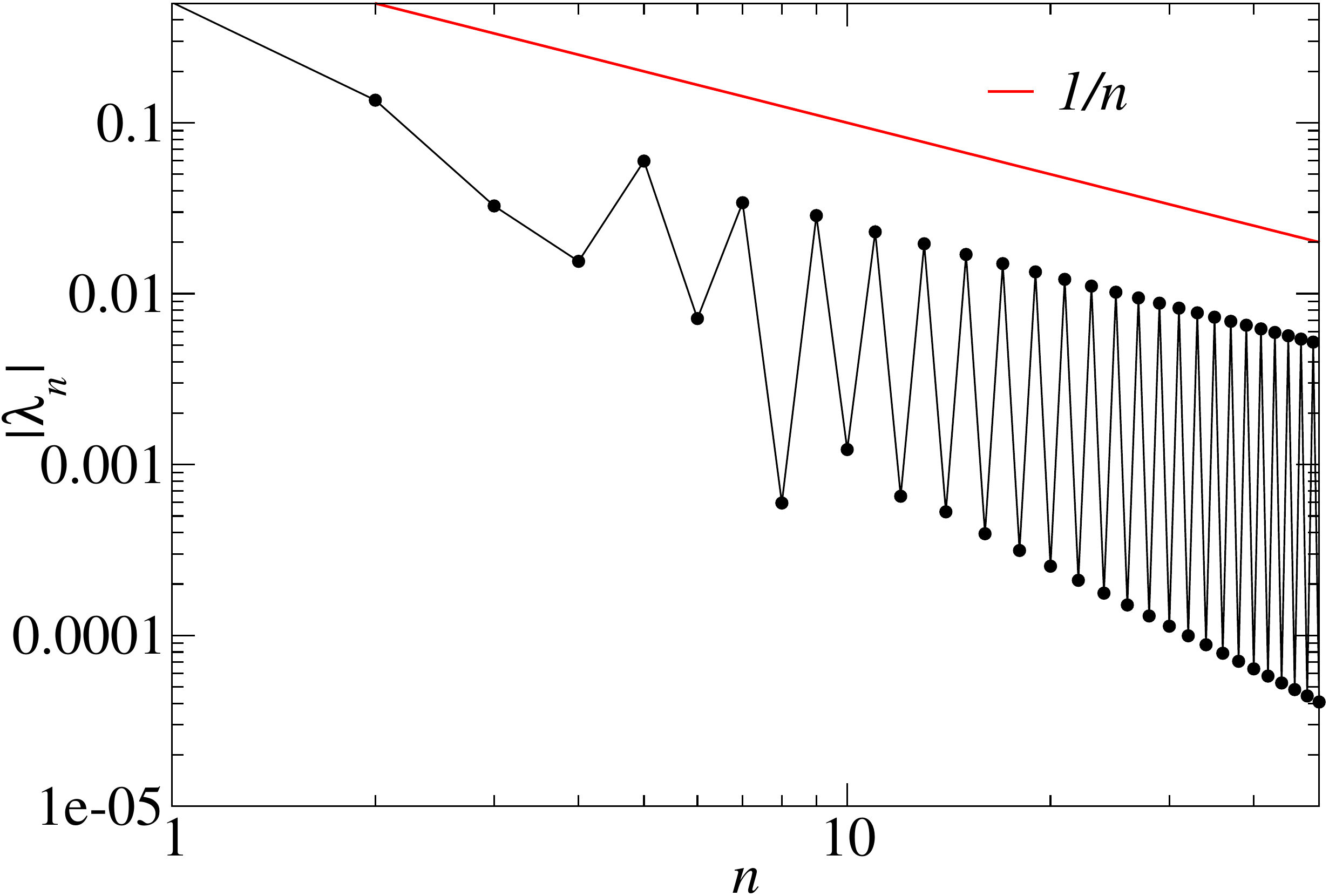}}\\
{\includegraphics[width=\hsize]{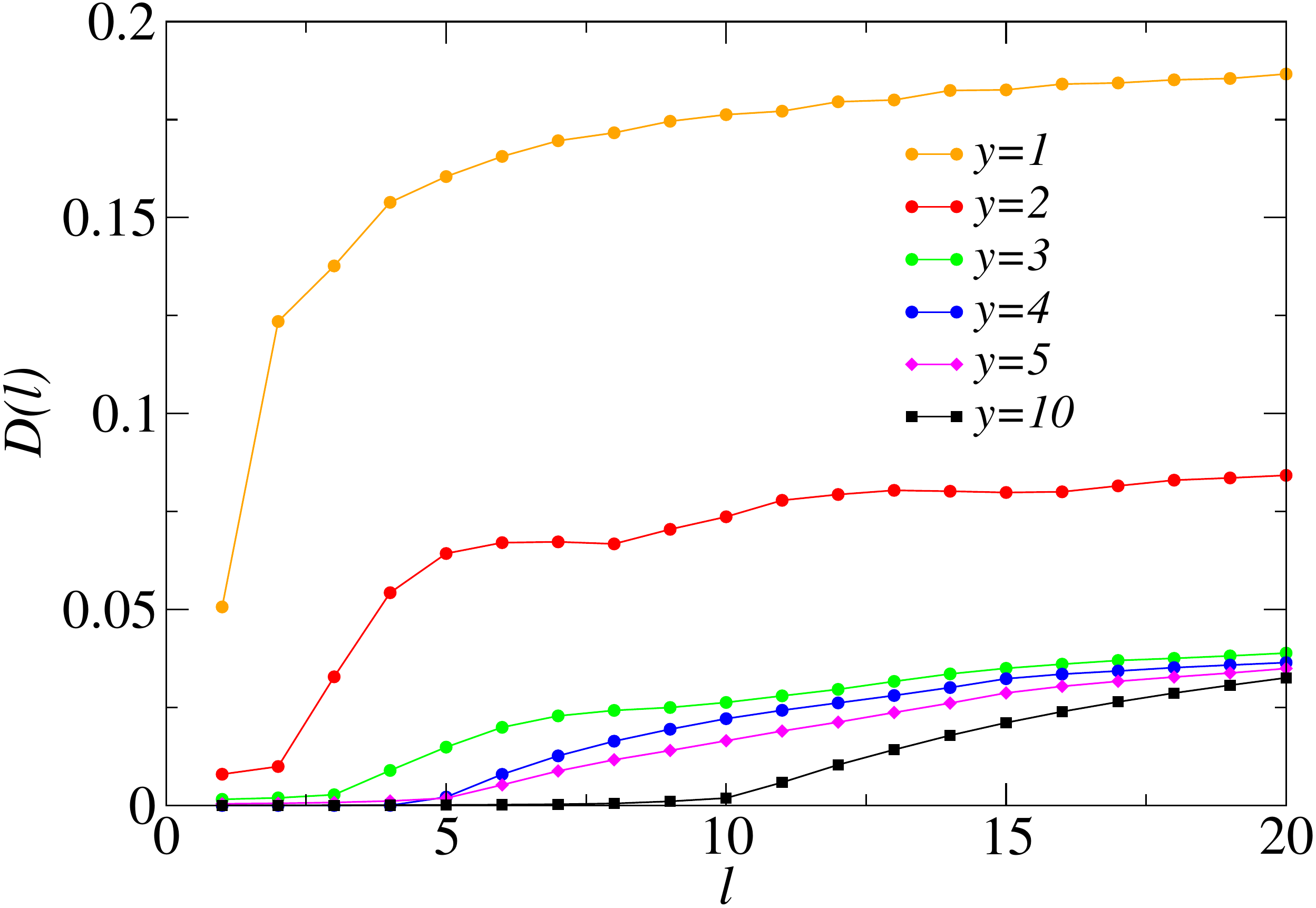}}
\caption{ (Top panel) The decay of $|\lambda_n|$ as a function of $n$ in an
  eigenstate where the coarse-grained $n_c(k)$ is discontinuous in the
  thermodynamic limit with $p_1 = 0.1$ and $p_2=0.4$. (Bottom panel) Results
  for the distance measure $D(l)$ for such a highly atypical eigenstate
  generated with 
$p_1 = 0.1$ and $p_2=0.4$ for a chain size of $L=2000$ as a function of $l$ with different truncated GGEs. 
\label{paperfig8}}
\end{figure}

For typical eigenstates drawn from generalized Microcanonical ensembles in the
TFIM, we have demonstrated that only a {\it few} Lagrange multipliers are
necessary for describing all the local properties of the state in a
thermodynamically large system and all other Lagrange multipliers can be set
to zero. E.g. a GE with only $\lambda_0=\beta$ being non-zero, and all other
Lagrange multipliers $\lambda_n=0$, provides the description for all local
properties for typical eigenstates drawn from a Microcanonical ensemble as
 shown in Sec.~\ref{results_typical}. What about eigenstates where the
 Lagrange multipliers $\lambda_n$ {\it cannot} be set to be zero beyond a
 certain $n$ and a full GGE description is therefore necessary? These
 eigenstates nonetheless have a finite energy density $E/L$ and show a volume
 law behaviour for the entanglement entropy and we study the RDMs of these
 states in this section. These eigenstates are generated by ensuring that the
 coarse-grained (in momentum space) Bogoluibov fermion occupation $n_c(k)$ is
 a {\it discontinuous} function of $k$ in the thermodynamic limit. There are
 several ways to achieve this and most simply, these eigenstates are obtained
 by placing the Bogoluibov fermion occupations $n_k=0 (1)$ with probability
 $1-p_1 (p_1)$ in the first $L/2$ positive $k$ modes and with a different
 probability $1-p_2 (p_2)$ in the next $L/2$ modes. For a large chain size
 $L$, any typical realization of this random process of placing $n_k=0(1)$
 generates an energy eigenstate where the coarse-grained $n_c(k)$ is $p_1$ for
 $0 \leq k \leq \pi/2$, and $p_2$ for $\pi/2 \leq k \leq \pi$. The
 discontinuity in $n_c(k)$ then leads to a slow decay of $|\lambda_n|$ as
 $1/n$ (see Fig.~\ref{paperfig8} (Top panel)) for any $p_1 \neq p_2$. In Fig.~\ref{paperfig8} (Bottom panel), we
 show the results of the comparison of the RDMs
 for such an atypical eigenstate in a chain size of $L=2000$, where we take
 $p_1 = 0.1$ and $p_2=0.4$, with
 various truncated GGEs.

Even though the full GGE is required if we need the accurate description of {\it all} local properties for such eigenstates in the thermodynamic limit, we clearly see that depending on the subsystem size $l$ being considered, one still requires only the first $y \sim l$ most local conservation laws for an ``accurate'' description of the properties of even such eigenstates and not {\it all} the integrals of motion from the behaviour of the distance measure $D(l)$ (Fig.~\ref{paperfig8}, bottom  panel). Going to bigger subsystems requires specifying a larger number ($y$) of integrals of motion to reduce $D(l)$. However, even for such eigenstates, we see that the most local integrals of motion play the most important role in describing local properties, and this constitutes the ``eigenstate thermalization version'' of a similar conclusion reached in Ref.~\onlinecite{Fagotti_Essler_PRB} for steady states following quantum quenches in the $1$D TFIM.        
  
\section{Quench from a typical eigenstate}
\label{quench}

We now address the situation when the Hamiltonian is time-dependent and consider the simplest case of a quantum quench, where the magnetic field ($g$) is suddenly changed from a pre-quench to a post-quench value at $t=0$. Typically, when considering quantum quenches, the starting state is assumed to be the ground state of the pre-quench Hamiltonian. Here, we consider the case where the initial state is instead a typical excited eigenstate of the pre-quench Hamiltonian (see also Ref.~\onlinecite{Calabrese_quench_excited_state}). Since such initial states are locally thermal as we explictly showed in the previous sections, a natural question is whether the unitary dynamics following the quantum quench at $t=0$ keeps them thermal at long times. 

We will show here that this is not the case since such states do not
have a finite overlap with the typical eigenstates of the post-quench
Hamiltonian but only with its rare eigenstates in the thermodynamic
limit. However, if the initial state was an
eigenstate of a non-integrable model, the final steady state 
might appear thermal~\cite{Rigol_PRL_quench}.
Thus, the long time description of the steady state again requires a GGE but
there are important differences 
when compared to a ground state quench.
\begin{figure}[htb]
{\includegraphics[width=\hsize]{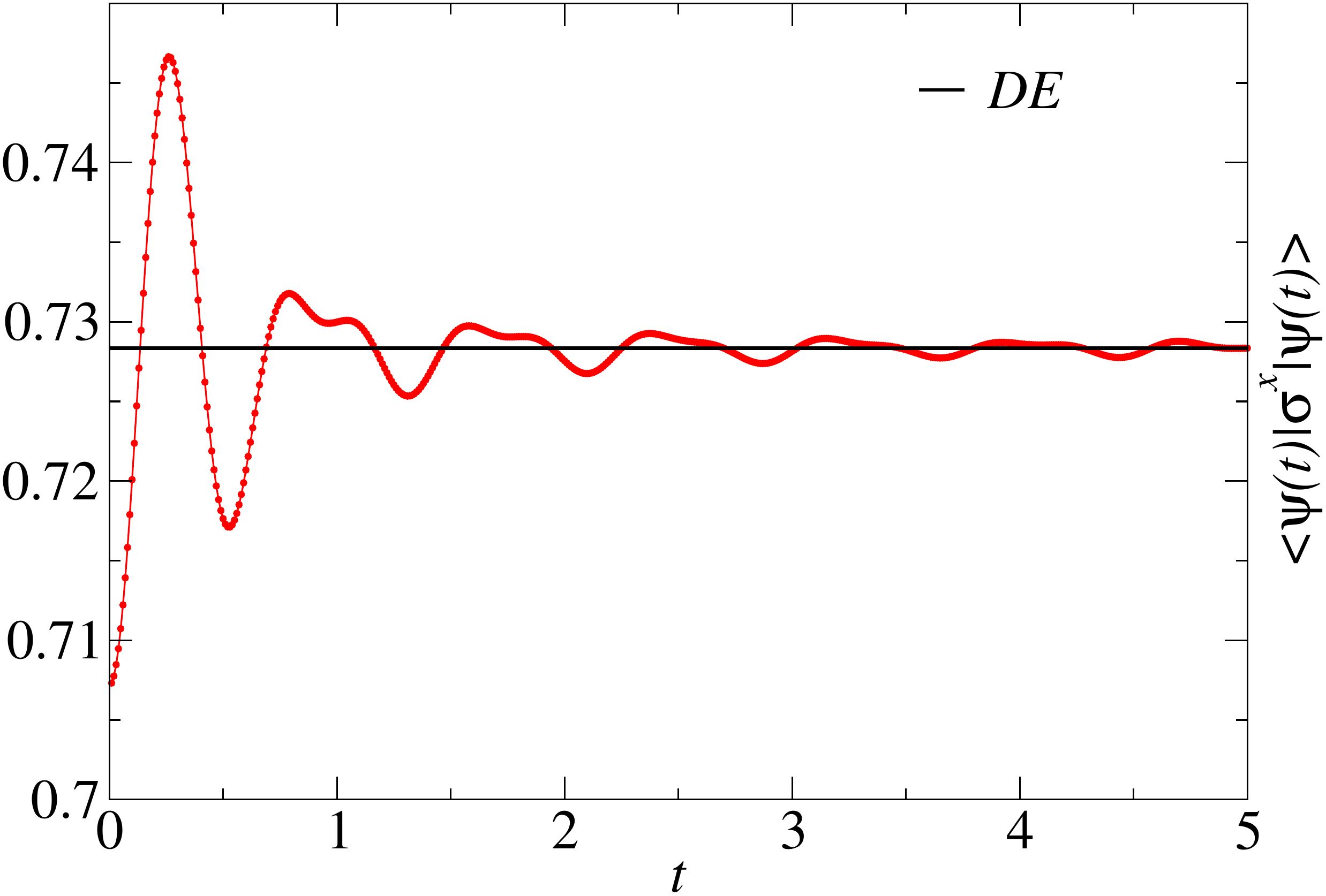}}
\caption{ The time evolution of the expectation value of the local operator $\sigma^x$ where the starting state at $t=0$ is a typical eigenstate (generated from the demon algorithm) at the coupling $g_i=2$ and chain size $L=30000$. The system's magnetic field is quenched to $g_f=3$ at $t=0$ and the unitary dynamics of the state leads to the Diagonal ensemble at long time (but smaller than the revival time of the finite system).  
\label{paperfig9}}
\end{figure}

The density matrix of the full system at a finite time can be written formally as 
\be
&&\rho(t) = |\psi(t) \rangle \langle \psi(t)| \nonumber \\
&=&   \sum_i |\langle \psi_{i}^f|\psi (0) \rangle|^2 | \psi_{i}^f \rangle \langle   \psi_{i}^f | \nonumber \\
&+& \sum_{i_1 \neq i_2}  e^{-i(E_{i_1}^f-E_{i_2}^f)t}\langle \psi_{i_1}^f| \psi(0) \rangle \langle \psi(0)| \psi_{i_2}^f \rangle  | \psi_{i_1}^f \rangle \langle   \psi_{i_2}^f | \nonumber \\
\ee
where $|\psi_{i}^f \rangle$ represent the eigenstates of the post-quench
Hamiltonian, $E^f_i$ its energy and $|\psi(0) \rangle$ denotes the starting
state at $t=0$. 
In the thermodynamic limit, the $i_1 \neq i_2$ terms cancel each other out
when $t \rightarrow \infty$~\cite{Rigol_nature, Caux_Essler_PRL} 
and hence, the density matrix of the steady state of the system coincides with the Diagonal ensemble (DE), described by the time-independent density matrix $\rho_{\rm{DE}}$:
\be
 \rho_{\rm{DE}} = \sum_i |\langle \psi_{i}^f|\psi(0) \rangle|^2 | \psi_{i}^f \rangle \langle   \psi_{i}^f |.
\ee 
In a finite system, any local operator will show revivals but this time scale
becomes progressively larger and diverges~\cite{Quantumrevivals_PRA} as $L \rightarrow \infty$.

The time-dependent wavefunction $|\psi (t) \rangle = \otimes_{k>0}|\psi_k (t)\rangle$ for the quench, where $|\psi_k (t)\rangle = U^n_{k}(t)c^{\dagger}_k c^{\dagger}_{-k} |0 \rangle + V^n_{k}(t)|0 \rangle$, can be easily worked out for $t>0$, by expressing the $t=0^-$ state at each $k$ in terms of the $(U_{k0(1)}(g_f),V_{k0(1)}(g_f))$ at the new coupling $g_f$. Doing this, we obtain $U^n_k(t)$ and $V^n_k(t)$ as follows:
\begin{widetext}
\be
 \left( \begin{array}{c} U^0_{k}(t) \\ V^0_{k}(t) \end{array} \right) &=& +e^{+i\epsilon_k(g_f)t} \cos ((\theta^i_k-\theta^f_k)/2) \left( \begin{array}{c} U_{k0}(g_f) \\ V_{k0}(g_f) \end{array} \right) + e^{-i\epsilon_k(g_f)t} \sin ((\theta^i_k-\theta^f_k)/2) \left( \begin{array}{c} U_{k1}(g_f) \\ V_{k1}(g_f) \end{array} \right) \nonumber \\
\left( \begin{array}{c} U^1_{k}(t) \\ V^1_{k}(t) \end{array} \right) &=& -e^{+i\epsilon_k(g_f)t} \sin ((\theta^i_k-\theta^f_k)/2) \left( \begin{array}{c} U_{k0}(g_f) \\ V_{k0}(g_f) \end{array} \right) + e^{-i\epsilon_k(g_f)t} \cos ((\theta^i_k-\theta^f_k)/2) \left( \begin{array}{c} U_{k1}(g_f) \\ V_{k1}(g_f) \end{array} \right)
\ee
\label{timedependence}
\end{widetext}
with the $n=0(1)$ index in $U^n_k(t)$ and $V^n_k(t)$ denoting whether the $t=0^-$ eigenstate at $g=g_i$ is $(U_{kn}(g_i),V_{kn}(g_i))$ at momentum $k$, and $\theta_k^i,\theta_k^f$ denote the Bogoluibov angles $\theta^g_k$ (see Eq.~\ref{eigenstates1}) for the pre-quench ($g_i$) and post-quench ($g_f$) values of the magnetic field respectively. We show the result of $\langle \psi(t)| \sigma^x | \psi(t) \rangle$ for a typical eigenstate at $g_i=2$ generated from the demon algorithm with $e_T = 0.3986$ where the magnetic field is quenched to $g_f=3$ at $t=0$ in Fig.~\ref{paperfig9}. We see that that $\langle \psi(t)| \sigma^x | \psi(t) \rangle$ already converges close to the DE result after a relatively short time $t \sim 5$. 

\begin{figure}[htb]
{\includegraphics[width=\hsize]{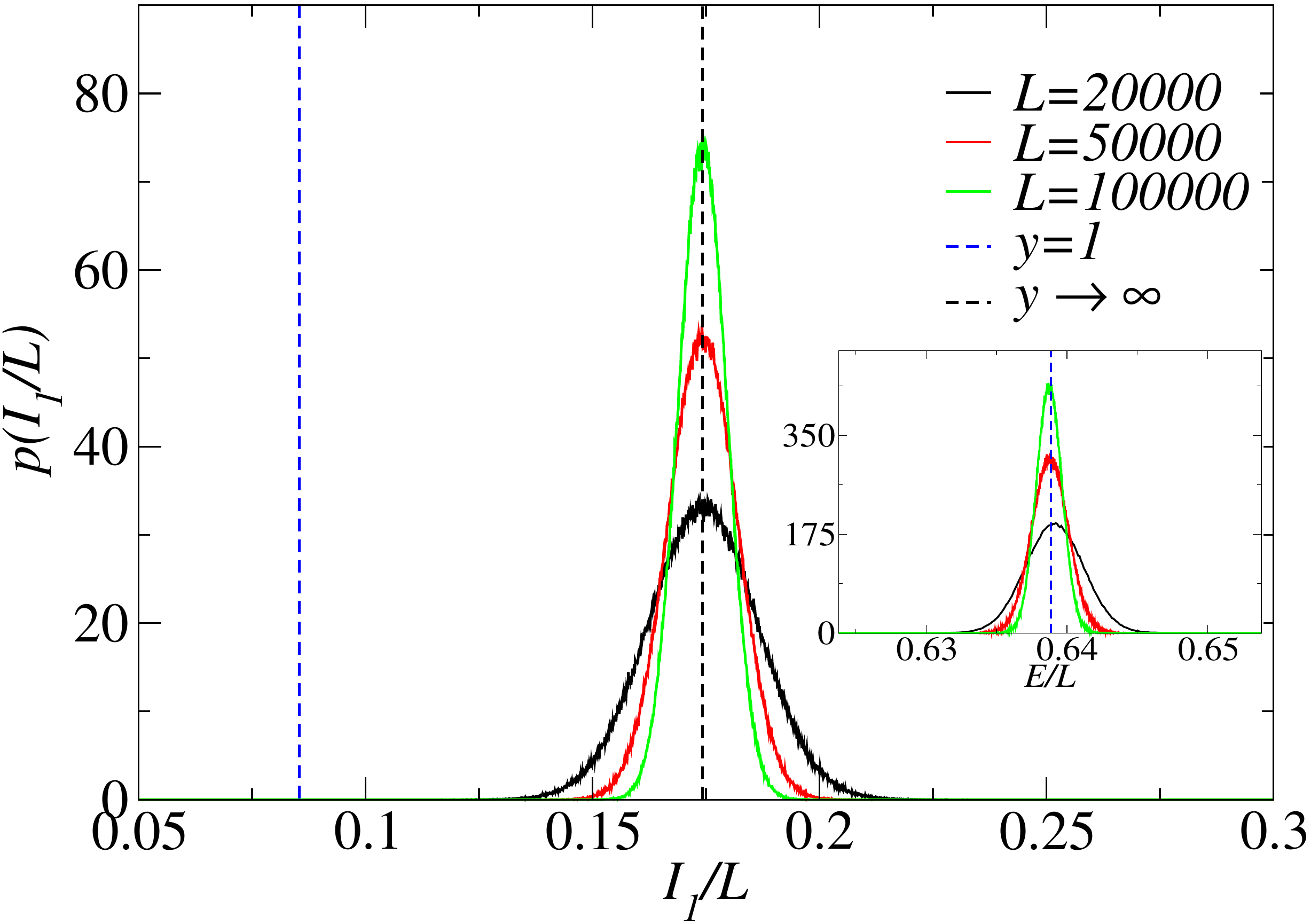}}\\
{\includegraphics[width=\hsize]{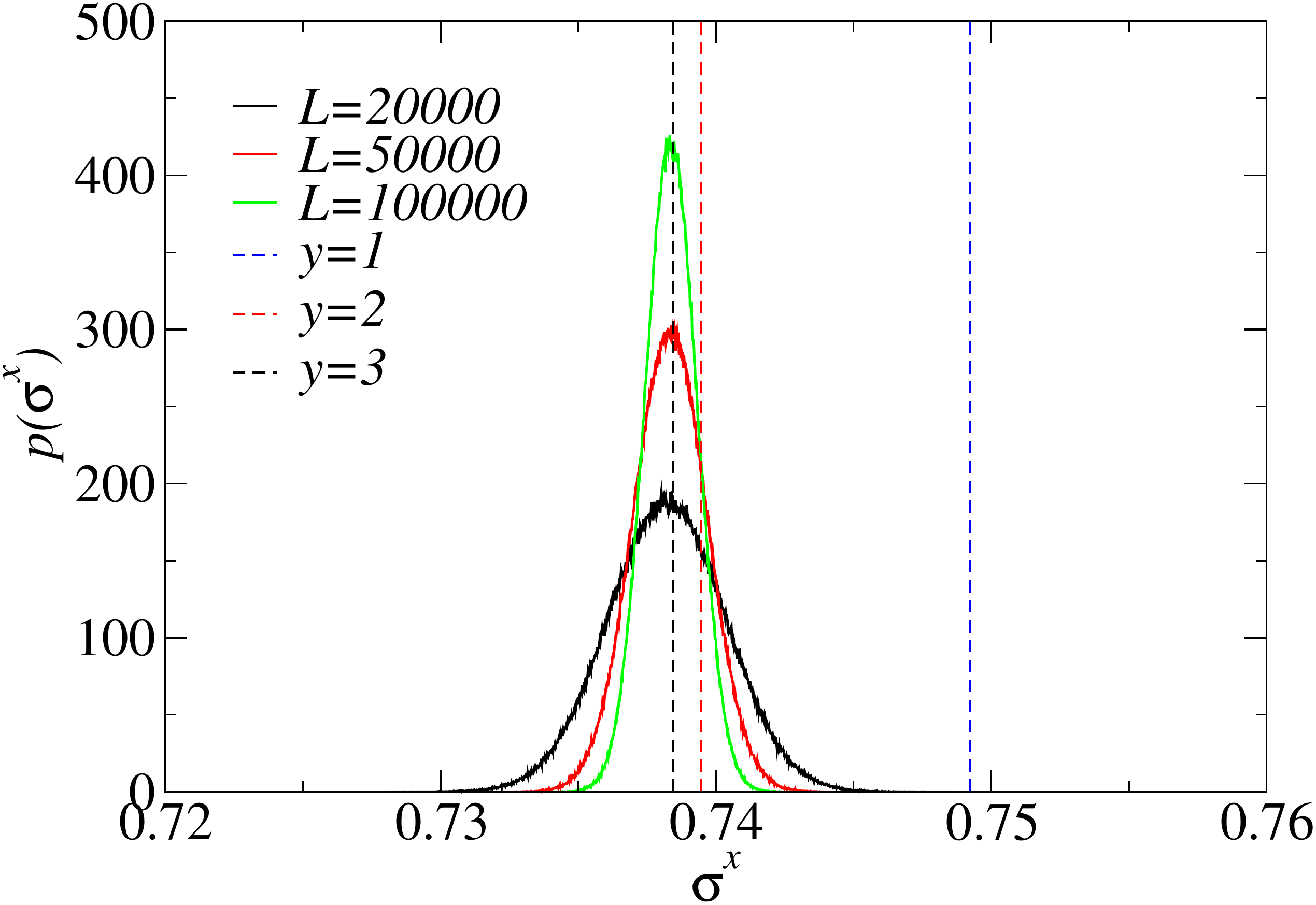}} 
\caption{ (Top panel) The probability density of $\langle I_1(g_f) \rangle/L$
  obtained from the steady state of the system starting from typical
  eigenstates at chain size $L$ and at coupling $g_i=2$ and quenched to a
  post-quench coupling of $g_f=3$. The inset shows the corresponding
  probability density for $\langle H(g_f)
\rangle /L$ in the steady state. (Bottom panel) The corresponding probability
density of $ \langle \sigma^x \rangle$ in the steady state. 
Clearly, the system's steady state is no longer described by a GE after the quench.    
\label{paperfig10}}
\end{figure}

We next calculate the steady state values of local observables $\langle H(g_f)
\rangle /L$, $\langle I_1(g_f) \rangle/L$ and $ \langle \sigma^x \rangle$
given that the quench starts from each of the sampled eigenstates from the MC
(with $g_i=2$ and $e_T = 0.3986$) at $t=0$  using the appropriate DE
determined by the initial state at $g_i$ and the value of $g_f$ and 
show the probability distributions of these steady state quantities in Fig.~\ref{paperfig10}. 

The distribitions of the DE values for the quantities shown in
Fig.~\ref{paperfig10} are normally distributed and the standard deviation
progressively shrinks to zero as $L$ is increased, which implies that in the
thermodynamic limit, quenches originating from {\it any} typical eigenstate
characterised by the same initial energy density $e_T$ at coupling $g_i$ lead
to steady states which are identical as far as local properties are
concerned. However, the mean values of the steady state distributions of
$\langle I_1(g_f) \rangle/L$ (Fig.~\ref{paperfig10}, Top panel)  and $ \langle
\sigma^x \rangle$ (Fig.~\ref{paperfig10}, Bottom panel) are very different
from the expected GE results fixed by the
 mean energy density of the final post-quench Hamiltonian (see inset of Top
 panel in Fig.~\ref{paperfig10}). 
Thus the steady state obtained after a quantum quench from a typical
eigenstate of the pre-quench Hamiltonian is not thermal and 
further conservation laws are needed for a quantitative agreement.

We now detail the construction of the GGE in the thermodynamic limit, and give the analytic expression for the Lagrange multipliers $\lambda_n$. The mean of the probability distributions of the different quantities shown in Fig.~\ref{paperfig10} ($\langle H(g_f) \rangle /L$, $\langle I_1(g_f) \rangle/L$ and $ \langle \sigma^x \rangle$) are all correctly captured by this GGE, and provides strong numerical support for its correctness in the thermodynamic limit. 

After a quench, the average Bologuibov fermion number $\langle n_k^{i,f} \rangle$ at each $k$ is conserved (and thus does not change as a function of $t$) because of the form of the post-quench Hamiltonian. Then, we have 
\be
\langle n_k^{i,f} \rangle = (p)\sin^2 \left(\frac{\theta_k^i -\theta_k^f}{2}  \right)+(1-p)\cos^2 \left(\frac{\theta_k^i -\theta_k^f}{2}  \right) \nonumber \\
\label{athermal_n}
\ee 
where $p=0(1)$ if $n_k(g_i) = 0(1)$ for the eigenstate of $g_i$ at $t=0$. In the thermodynamic limit, all the microscopic $n_k(g_i)$ lead to the same coarse-grained $n_c(k)$ which follows a thermal distribution that is fixed only by the average energy density of the eigenstate. Thus, when $L \rightarrow \infty$, we can replace the $p$ variables (which equal $n_k(g_i)$ microscopically) by the same thermal distribution to get its coarse-grained version:
\be
p_c(k) = \frac{\exp(-\beta \epsilon_k^i)}{\exp(-\beta \epsilon_k^i)+\exp(+\beta \epsilon_k^i)}
\ee
where $\beta$ is the inverse temperature of the GE that describes the local properties of the typical eigenstates at $g_i$. Thus, in the $L \rightarrow \infty$ limit, we obtain
\be
\langle n_k^{i,f} \rangle = \frac{e^{-\beta \epsilon_k^i} \cos^2 \left(\frac{\theta_k^i -\theta_k^f}{2}\right) +e^{\beta \epsilon_k^i} \sin^2 \left(\frac{\theta_k^i -\theta_k^f}{2}  \right)}{e^{-\beta \epsilon_k^i} + e^{\beta \epsilon_k^i}}
\ee

The Lagrange multipliers $\lambda_n$ wrt the final post-quench Hamiltonian (at $g_f$) are then defined by using Eqn.~\ref{lambdak} and Eqn.~\ref{lambdan}:
\be
\lambda_n &=&  \frac{2-\delta_{n,0}}{\pi}\int_0^\pi \cos(nk) \mathcal{F} (g_i,g_f,\beta,k) \nonumber \\
\mathcal{F} (g_i,g_f,\beta,k) &=& \frac{1}{2\epsilon^f_k} \log \left(\frac{1-\langle n_k^{i,f} \rangle}{\langle n_k^{i,f} \rangle} \right)
\ee

Thus, knowing the initial energy density of the typical eigenstate at the pre-quench magnetic field value of $g_i$, and the couplings $g_i$ and $g_f$, completely fixes the Lagrange multipliers ($\lambda_n$) and hence the GGE from Eqn.~\ref{GGE_y}. The values obtained from this GGE are fully consistent with the mean values of $\langle H(g_f)\rangle/L, \langle I_1(g_f)\rangle/L, \sigma^x$ in the steady state around which the standard deviation shrinks to zero as $L \rightarrow \infty$ in Fig.~\ref{paperfig10}. At low $\beta$, this expression can be further simplied to give 
\be
\lambda_n = \left(\frac{2-\delta_{n,0}}{\pi} \right) \beta  \int_0^\pi \left( \frac{\epsilon^i_k}{\epsilon^f_k} \cos(\theta_k^i -\theta_k^f) \right) \cos(nk)
\ee 
Clearly, only when $\beta = 0$ for the initial pre-quench eigenstate is the
final steady state also thermal (with $\beta=0$ again) with respect to the
final post-quench Hamiltonian. Even at small $\beta$, $\lambda_n$ for $n>0$
are non-zero (though small) and hence one obtains a GGE for the steady
state. The athermal nature of the ensemble is related to the athermal
behaviour of $\langle n_k^{i,f} \rangle$ (Eqn.~\ref{athermal_n}), which fixes
all the (local) conserved quantities, since it cannot be expressed as
$\exp(-\beta_f \epsilon_k(g_f))/(\exp(+\beta_f \epsilon_k(g_f))+\exp(-\beta_f
\epsilon_k(g_f)))$ for any $\beta_f$ as long as the initial $\beta \neq 0$.

 \begin{figure}[htb]
{\includegraphics[width=\hsize]{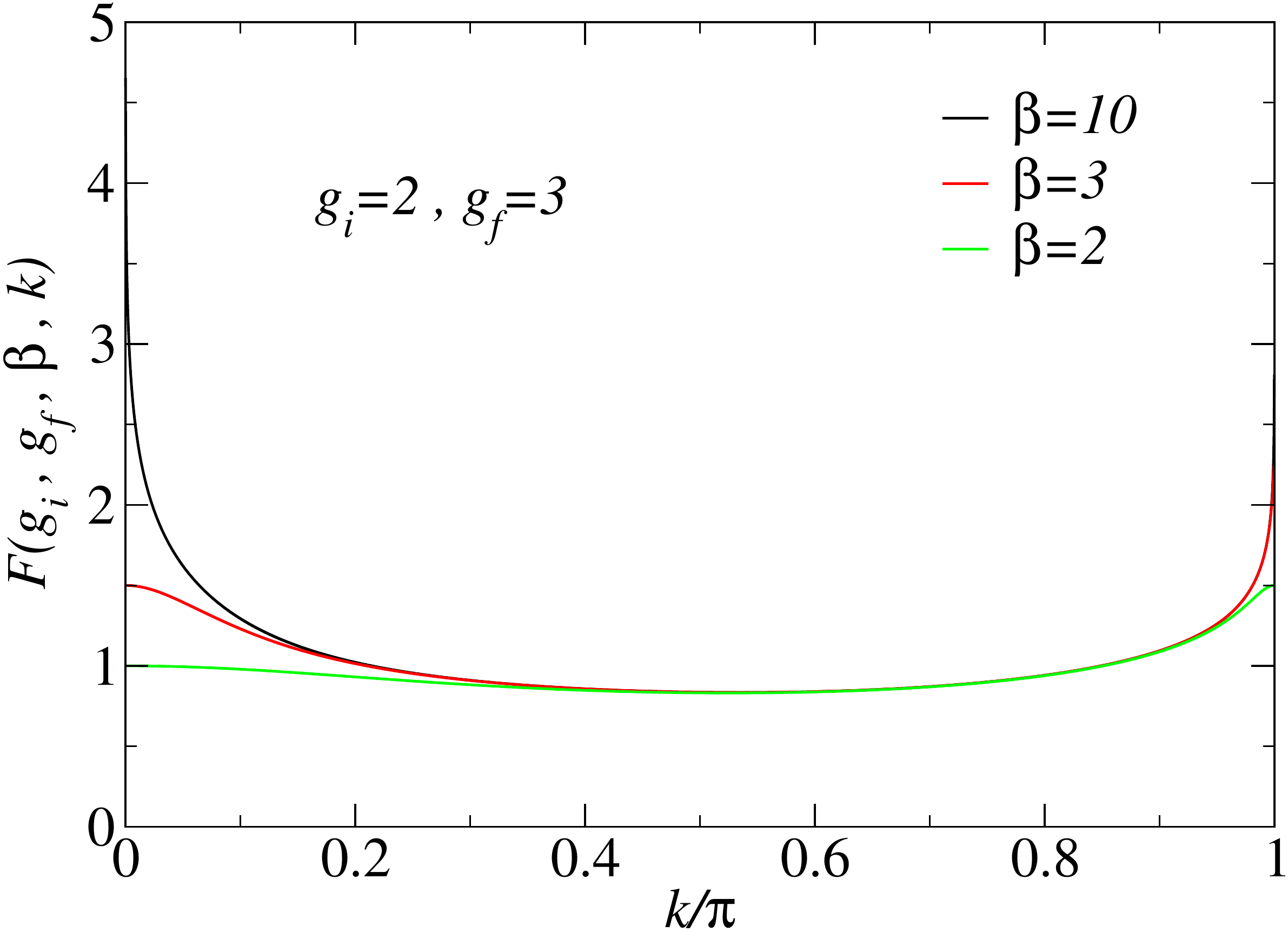}} 
{\includegraphics[width=\hsize]{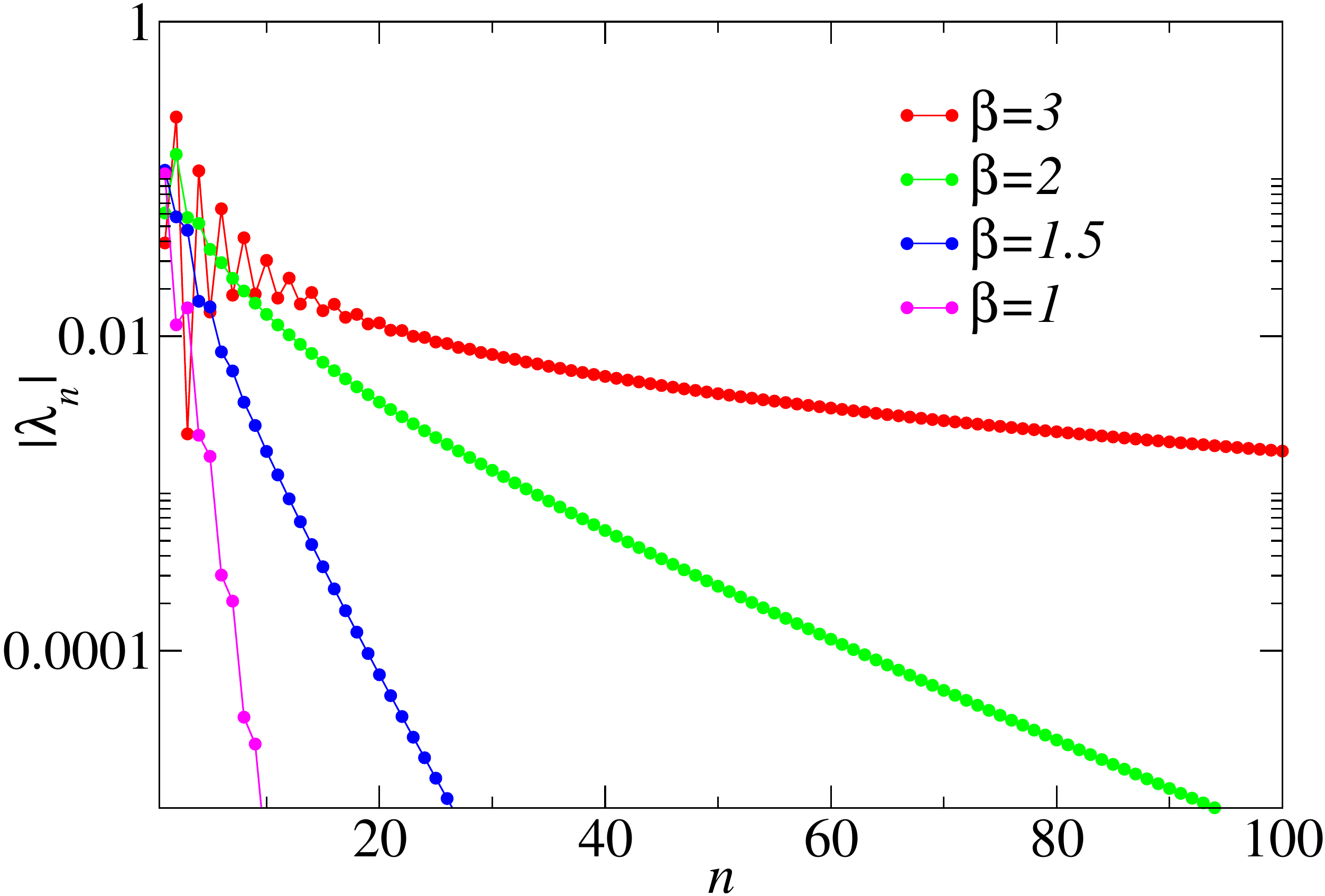}}
\caption{(Top panel) The singularities present in $\mathcal{F}
  (g_i,g_f,\beta \rightarrow \infty,k)$ at $k=0,\pi$ get rounded off at finite
  $\beta$ that corresponds to the energy density of a typical (pre-quench
  Hamiltonian's) eigenstate. (Bottom panel) 
The decay of the Lagrange multipliers for the post-quench GGE as a function of
distance $n$. 
Here, $g_i=2$, $g_f=3$ and $\beta$ is fixed only by the energy density of the typical eigenstate of the pre-quench Hamiltonian.
\label{paperfig11}}
\end{figure}

Note that the $\lambda_n$ when the initial state is the ground state of the
pre-quench Hamiltonian can be simply obtained by taking $\beta \rightarrow
\infty$ and matches the results obtained in that context by Fagotti and
Essler~\cite{Fagotti_Essler_PRB}. From this previous work, it is known that
$\lambda_n$ decay rather slowly with distance as $1/n$ when the $t=0$ state is
the pre-quench Hamiltonian's ground state, because of the logarithmic
singularity of $\mathcal{F}(g_i,g_f,\beta \rightarrow \infty,k)$ at $k=0$ and
$k=\pi$. 
However, for any finite $\beta$ (which corresponds to a highly excited
eigenstate at $t=0$), the singularities are rounded off as shown in
Fig.~\ref{paperfig11} (Top panel). 
This instead leads to an exponential decay of $|\lambda_n| \sim \exp(-n/\xi)$
as shown in Fig.~\ref{paperfig11} (Bottom panel), where $\xi$ indicates the
length-scale associated with the exponential decay in $n$. 
It will be interesting to obtain an analytic expression for $\xi$ as a function of $\beta,g_i,g_f$. 
 
\section{Conclusions}
\label{conclude}
We have studied the reduced density matrices and local properties of highly excited eigenstates of the transverse field Ising chain, sampling them using an unbiased Monte-Carlo technique. We find that, in spite of being integrable with an extensive number of conserved quantities, typical high energy eigenstates are described by a finite temperature Gibbs ensemble for all local properties in the thermodynamic limit. Our sampling method also allows us exploring rare (athermal) eigenstates, and we explictly demonstrate that such states are locally described by appropriate truncated Generalized Gibbs ensembles with only a few non-zero Lagrange multipliers. We also consider a class of high energy eigenstates for which the full GGE is required to describe local properties accurately. Nonetheless, the most local conservation laws still play the most important role in describing local properties. We, however, show that even for a quantum quench from a typical high-energy eigenstate of the pre-quench Hamiltonian, the resulting steady state requires a full GGE description. Our study leaves many open issues for future studies. For example, it will be interesting to investigate the behaviour of unequal time correlation functions of high energy excited states, especially in light of the results presented in Ref.~\onlinecite{Maldacenaetal_Chaos}. Another interesting question is whether this picture of typicality holds
for free Hamiltonians with long range interctions. A related question is regarding the typical nature of the periodic Gibbs' ensemble~\cite{AAR-PRL} produced by driving 
free-fermions (or other integrable models mapable to that) periodically: if we observe the asymptotic synchronized state stroboscopically, do we typically get a thermal 
state? The question is interesting, since the effective Floquet hamiltonian, though still bilinear in fermions, may be long-ranged, and can often be non-local in terms
of the original degrees of freedom.

\bibliographystyle{apsrev4-1}
\bibliography{Bib_Typicality}

\end{document}